\documentclass[journal]{IEEEtran}
\usepackage{amsmath}
\usepackage{amssymb}
\usepackage{amsfonts}
\usepackage{graphicx}
\usepackage{epsfig}
\usepackage{subfigure}
\usepackage{psfrag}
\usepackage{cite}
\usepackage{latexsym}
\usepackage{url}
\usepackage{color}
\usepackage{multirow}
\usepackage{bm}
\usepackage{multicol,comment}

\usepackage[colorlinks,linkcolor=black,anchorcolor=black,citecolor=black]{hyperref}
\begin{document}
\title{Cognitive UAV Communication via Joint Maneuver and Power Control}
\author{Yuwei~Huang,~Weidong~Mei,~Jie~Xu,~\IEEEmembership{Member,~IEEE,}~Ling~Qiu,~\IEEEmembership{Member,~IEEE,}

and~Rui Zhang,~\IEEEmembership{Fellow,~IEEE}\vspace{-10pt}
\thanks{

\scriptsize{This work has been presented in part at the IEEE International Workshop on Signal Processing Advances in Wireless Communications (SPAWC), June 25--28, 2018, Kalamata, Greece \cite{my}.}

\scriptsize{Y. Huang and L. Qiu are with the Key Laboratory of Wireless-Optical Communications, Chinese Academy of Sciences, School of Information Science and Technology, University of Science and Technology of China, Hefei, Anhui, 230027, China (e-mail: hyw1023@mail.ustc.edu.cn, lqiu@ustc.edu.cn).}

\scriptsize{W. Mei is with the NUS Graduate School for Integrative Sciences and Engineering, National University of Singapore, and also with the Department of Electrical and Computer Engineering, National University of Singapore (e-mail: wmei@u.nus.edu).}

\scriptsize{J. Xu is with the School of Information Engineering, Guangdong University of Technology, and also with the National Mobile Communications Research Laboratory, Southeast University (e-mail: jiexu@gdut.edu.cn). J. Xu is the corresponding author.}

\scriptsize{R. Zhang is  with the Department of Electrical and Computer Engineering, National University of Singapore (e-mail: elezhang@nus.edu.sg).}
}}
\maketitle
\newcommand{\mv}[1]{\mbox{\boldmath{$ #1 $}}}
\newtheorem{lemma}{\underline{Lemma}}[section]
\newtheorem{lemma1}{\underline{Lemma}}
\newtheorem{remark}{\underline{Remark}}[section]
\newtheorem{proposition}{\underline{Proposition}}[section]
\newtheorem{corollary}{\underline{Corollary}}[section]

\begin{abstract}
This paper investigates a new scenario of spectrum sharing between unmanned aerial vehicle (UAV) and terrestrial wireless communication, in which a cognitive/secondary UAV transmitter communicates with a ground secondary receiver (SR), in the presence of a number of primary terrestrial communication links that operate over the same frequency band. We exploit the UAV's mobility in three-dimensional (3D) space to improve its cognitive communication performance while controlling the co-channel interference at the primary receivers (PRs), such that the received interference power at each PR is below a prescribed threshold termed as interference temperature (IT). First, we consider the quasi-stationary UAV scenario, where the UAV is placed at a static location during each communication period of interest. In this case, we jointly optimize the UAV's 3D placement and power control to maximize the SR's achievable rate, subject to the UAV's altitude and transmit power constraints, as well as a set of IT constraints at the PRs to protect their communications. Next, we consider the mobile UAV scenario, in which the UAV is dispatched to fly from an initial location to a final location within a given task period. We propose an efficient algorithm to maximize the SR's average achievable rate over this period by jointly optimizing the UAV's 3D trajectory and power control, subject to the additional constraints on UAV's maximum flying speed and initial/final locations. Finally, numerical results are provided to evaluate the performance of the proposed designs for different scenarios, as compared to various benchmark schemes. It is shown that in the quasi-stationary scenario the UAV should be placed at its minimum altitude while in the mobile scenario the UAV should adjust its altitude along with horizontal trajectory, so as to maximize the SR's achievable rate in both scenarios.
\end{abstract}
\begin{IEEEkeywords}
UAV communication, spectrum sharing, 3D placement, 3D trajectory design, power control, interference management.
\end{IEEEkeywords}

\section{Introduction}
With continuous technology advancement and cost reduction, unmanned aerial vehicles (UAVs) or drones have been more widely used in various applications, such as cargo delivery, aerial photography, surveillance, search and rescue, etc \cite{RZhangUAVchallenges}. It is projected by Federal Aviation Administration (FAA) \cite{FAA} that there will be around seven million UAVs in the United States only in 2020. With the explosively increasing number of UAVs, how to integrate them into future wireless networks to enable their bidirectional communications with the ground users/pilots has become a critical task to be tackled. On one hand, for emergency situations (e.g., after natural disaster) and temporary hotspots (e.g., stadium during a football match), UAVs can be employed as {\it aerial wireless communication platforms} (e.g., relays or base stations (BSs)) to provide data access, enhance coverage, and improve communication rates for ground users \cite{RZhangrelay,JXubroadcast}. On  the other hand, for UAVs in various missions (e.g., cargo delivery), it is crucial to enable them as {\it aerial mobile users} to access existing wireless networks (e.g., cellular networks), in order to support not only secure, reliable, and low-latency remote command and control, but also high-capacity mission-related data transmission \cite{shuowen,Yongcellular,Ismail}. Therefore, UAV-assisted terrestrial communications \cite{RZhangrelay} and network-connected UAV communications \cite{shuowen} have become two widely investigated paradigms for integrating UAVs into future wireless communication networks.

UAV communications are different from conventional terrestrial wireless communications in the following two main aspects. First, UAVs normally have strong line-of-sight (LoS) links with ground nodes, thus offering better channel conditions than terrestrial fading channels and even making it possible to predict channel state information (CSI) and hence communication performances at different UAV's three-dimensional (3D) locations based on the ground nodes' location information. Second, UAVs have fully controllable mobility in 3D, by exploiting which UAVs can adjust their altitude and horizontal location over time to optimize their communication performances with ground nodes.

In the literature, there are generally two lines of research that exploit the UAV mobility for communication performance optimization, namely the {\it quasi-stationary UAV} with 3D placement optimization and the {\it mobile UAV} with 3D trajectory optimization, respectively. For the quasi-stationary scenario, the UAV is placed at a static location over each communication period of interest, while its location can be changed from one period to another. This may practically correspond to a UAV communication platform that is connected by a cable/wire with a ground control platform (see, e.g., ``flying cell-on-wings (COWs)" of AT\&T  \cite{AT} and ``Air Masts" of Everything-Everywhere (EE: the UK's largest mobile network operator) \cite{EE}). Substantial research efforts have been devoted to this research paradigm. For example, the works \cite{staticaltitude} and \cite{staticaltitude2} optimized the UAV-BS's altitude to maximize the coverage probability on the ground and minimize the system outage probability, respectively;  while \cite{static3D2} optimized the UAV-BS's 3D location to maximize the number of served ground users. Furthermore, \cite{statichorizontal,multiple} and \cite{peiming} optimized multiple UAV-BSs'  locations to minimize the number of required UAV-BSs to cover a given area and maximize the minimum throughput among all ground users, respectively. In \cite{multiple2}, the downlink coverage probability for a reference ground user was analyzed in the presence of multiple UAV-BSs, while \cite{staticSE} showed that the deployment of UAV-BSs at their optimized locations can improve the coverage performance and spectral efficiency of the network. In addition, \cite{DGrelay} investigated the optimal placement of a UAV-relay to maximize the end-to-end throughput from a source to a destination by using a new LoS map based approach.

On the other hand, under the mobile UAV scenario, prior works have designed the UAV trajectory (i.e., 3D locations over time) jointly with communication scheduling and resource allocation for performance optimization. For example, when the UAV is employed as a mobile relay, the authors in \cite{RZhangrelay} and \cite{Shuhangrelay} optimized the UAV-relay's trajectory to maximize the end-to-end throughput and minimize the system outage probability, respectively. When UAVs are employed as cellular BSs, the authors in \cite{JXubroadcast,RZhangmulticast,Yundi,QWuscp,noma1} optimized the UAVs' trajectories to maximize the achievable rates under different setups such as broadcast channel \cite{JXubroadcast,noma1}, multicast channel \cite{RZhangmulticast,Yundi}, and interference channel \cite{QWuscp}. Furthermore, the UAV trajectory design was also investigated for other applications when the UAV is employed as an access point (AP) for wireless power transfer \cite{JXuWPT}, wireless powered communication \cite{JXu2018}, and mobile edge computing \cite{MEC}. In addition, when UAVs act as cellular users that perform tasks in a long range, the works \cite{shuowen} and \cite{Ismail} studied the UAV users' trajectory design to minimize the mission completion time, subject to various communication connectivity constraints with ground BSs. In \cite{Saad}, an interference-aware path planning design was proposed for multiple UAV users, which aimed to achieve an optimal trade-off between energy efficiency, latency and interference caused by the UAVs to the ground network. The work \cite{Saad2} proposed an energy-efficient path planning design to minimize the energy consumption of UAV swarms, subject to individual energy availability constraints at UAVs.

\begin{figure*}
\subfigure[Network-connected UAV communication.]{\label{cog_uav_uplink}
\begin{minipage}[t]{0.48\linewidth}
\centering
\includegraphics[width=7cm]{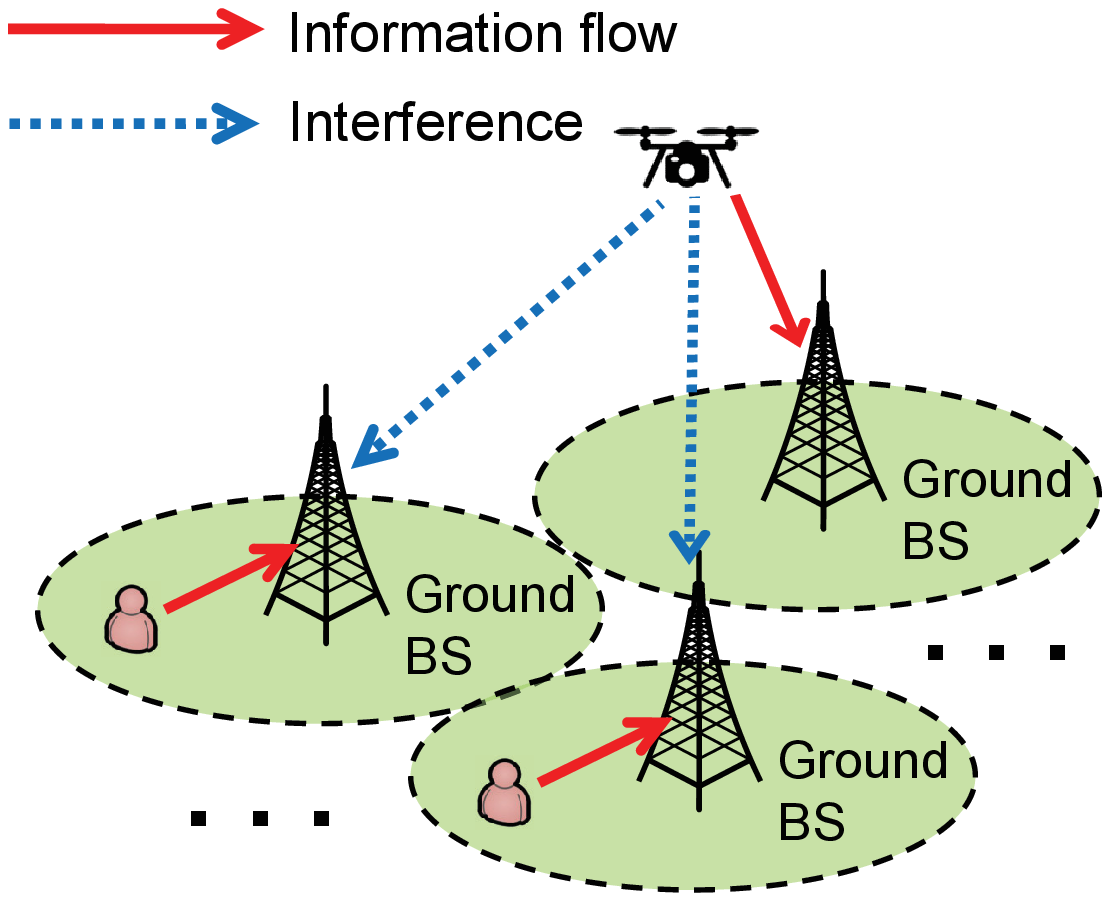}
\end{minipage}
}
\hfill
\subfigure[D2D-enabled UAV-ground communication.]{\label{cog_uav_D2D}
\begin{minipage}[t]{0.48\linewidth}
\centering
\includegraphics[width=7cm]{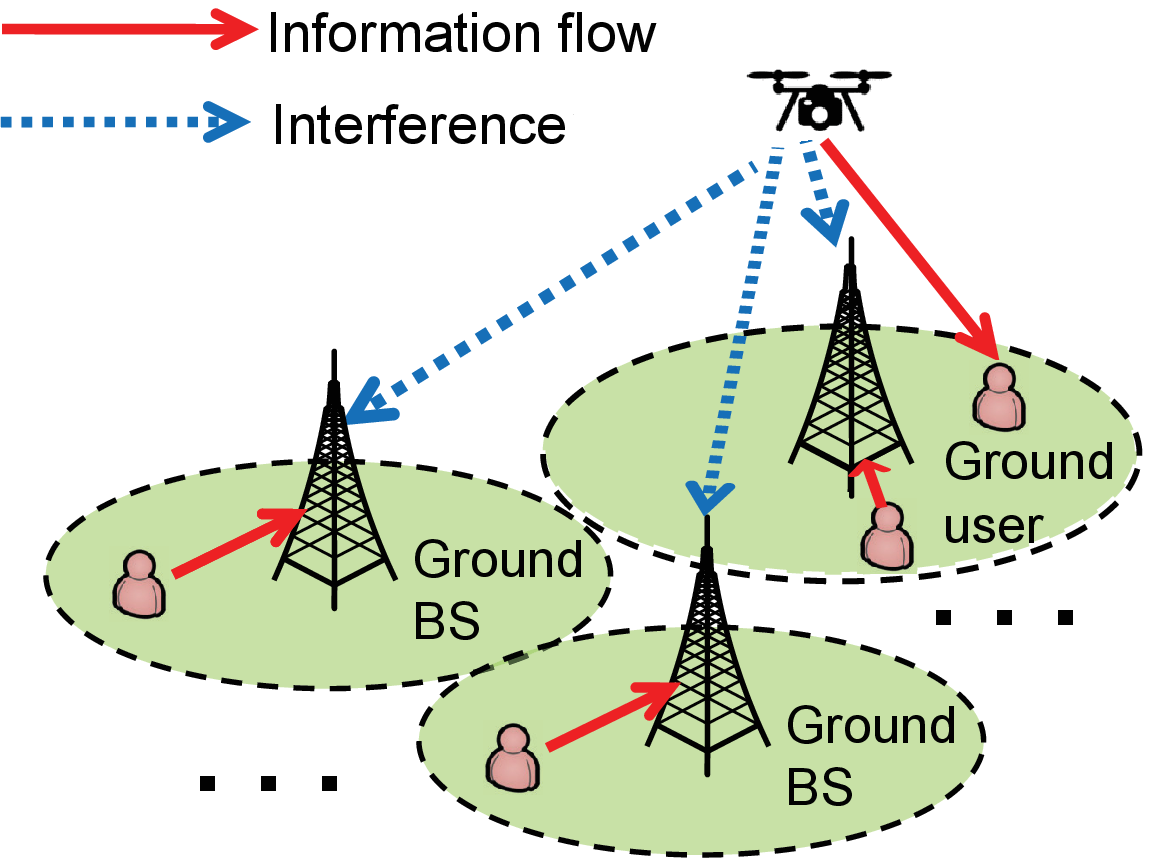}
\end{minipage}
}
\caption{Illustration of the cognitive UAV communication systems.}\label{cog_UAV_communication_picture}
\vspace{-10pt}\end{figure*}

Despite the above research progress, existing works have mostly assumed that the UAV communications are operated over dedicated frequency bands. Nevertheless, due to the scarcity of wireless  spectrum, it is practically difficult to allocate dedicated spectrum to new UAV communications. To address this challenge and motivated by the technical advancement of spectrum sharing in cognitive radio (CR) \cite{GoldSmith}, a viable solution is to allow UAVs to operate as cognitive or secondary communication nodes to access the spectrum that is originally allocated to existing (primary) terrestrial wireless communication networks (see, e.g., \cite{WZspectrumsharing}). For instance, in the network-connected UAV communication (as shown in Fig. \ref{cog_uav_uplink}), the UAV communicates with its associated ground BS by reusing the resource blocks (RBs)  assigned to existing ground users;
whereas in the device-to-device (D2D)-enabled UAV-ground communication (as shown in Fig. \ref{cog_uav_D2D}), the UAV communicates with its associated ground user via D2D communication by reusing the RBs in the uplink cellular communications. In both the above two cases, a new and severe air-to-ground (A2G) interference issue needs to be tackled \cite{Yongcellular,Polin,XLin,Mogensen}, since the A2G channels are normally LoS-dominated.  Specifically, the UAV may impose severe uplink interference to multiple co-channel non-associated ground BSs (primary receivers (PRs)) in network-connected UAV communication (Fig. \ref{cog_uav_uplink}). Similarly, the D2D communication from the UAV to ground user may impose severe uplink interference at co-channel ground BSs (PRs) in D2D-enabled UAV-ground communication (Fig. \ref{cog_uav_D2D}). As a result, how to maximize the UAV communication throughput while effectively mitigating the A2G co-channel interference to the primary communication system is an important and yet challenging problem that calls for innovative solutions. It is worth noting that there have been some initial studies on A2G interference mitigation for network-connected UAV communication in the literature \cite{weidong,weidong1,liuliang}, which, however, only considered the case of a static UAV user. By leveraging the UAV's controllable mobility, in this paper, we propose a new approach to tackle this problem, which jointly optimizes the UAV's 3D placement or trajectory (for the quasi-stationary and mobile UAV scenarios, respectively) and CR-based interference-aware transmit power control to achieve the maximum throughput of UAV-to-ground secondary communication, while controlling the interference to existing primary ground receivers below a tolerable level.

For the purpose of exposition, this paper considers a spectrum sharing system where a cognitive/secondary UAV transmitter communicates with a ground secondary receiver (SR), in the presence of a number of primary terrestrial communication links that operate over the same frequency band. Under this setup, we adopt the {\it interference temperature} (IT) technique in CR \cite{RZhangIT} to protect the primary communications, so that the received power at each PR cannot exceed a prescribed IT threshold. The main results of this paper are summarized as follows.
\begin{itemize}
\item First, we consider the quasi-stationary UAV scenario, in which the UAV is placed at an optimized location that is fixed during the communication period of interest. We jointly optimize the UAV's 3D placement and transmit power to maximize the SR's achievable rate, subject to the UAV's flight altitude and transmit power constraints, and a set of IT constraints at the PRs. The joint 3D placement and power optimization problem is non-convex and difficult to be optimally solved in general. To tackle this challenge, we first prove that the UAV should be placed at the lowest altitude at the optimality. Building upon this, we further use the semi-definite relaxation (SDR) technique to obtain the UAV's optimal horizontal location and transmit power.
\item Next, we consider the mobile UAV scenario, in which the UAV is dispatched to fly from an initial location to a final location during a particular task period. We maximize the SR's average achievable rate over this period by jointly optimizing the UAV's 3D trajectory and transmit power over time, subject to the UAV's maximum flying speed, altitude, and transmit power constraints, as well as the PRs' IT constraints. Due to the time-dependent UAV trajectory variables, this problem is more involved and thus more difficult to be solved as compared to that in the quasi-stationary scenario. To tackle this problem, we propose an efficient algorithm that ensures a locally optimal solution by applying the technique of successive convex approximation (SCA).
\item Finally, numerical results are presented to validate the performance of our proposed cognitive UAV communication designs, as compared to other benchmark schemes, for both the quasi-stationary and mobile scenarios. Specifically, it is shown that in the mobile scenario, the UAV needs to adaptively adjust its altitude together with horizontal location over time to balance the trade-off between maximizing the SR's rate versus minimizing the interference with PRs. This is in a sharp contrast to the quasi-stationary scenario, where it is shown that the UAV should always be placed at its lowest altitude at the optimality.
\end{itemize}

The remainder of this paper is organized as follows. Section $\text{\uppercase\expandafter{\romannumeral2}}$ introduces the system model of the cognitive UAV communication system and formulates the optimization problems of our interest. Section $\text{\uppercase\expandafter{\romannumeral3}}$ presents the optimal solution to the joint 3D placement and power control problem in the quasi-stationary UAV scenario. Section $\text{\uppercase\expandafter{\romannumeral4}}$ proposes an efficient algorithm to obtain a locally optimal solution to the joint 3D trajectory and power optimization problem in the mobile UAV scenario. Section $\text{\uppercase\expandafter{\romannumeral5}}$ provides numerical results to demonstrate the efficacy of our proposed designs versus benchmark schemes. Finally, Section $\text{\uppercase\expandafter{\romannumeral6}}$ concludes this paper.

{\it Notations:} In this paper, scalars are denoted by italic letters, vectors and matrices are denoted by bold-face lower-case and upper-case letters, respectively. $\mathbb{R}^{x\times y}$ denotes the space of $x\times y$ real-valued matrices. For a square matrix $\mv M$, $\text{Tr}(\mv M)$, $\text{det}(\mv M)$, and $\text{rank}(\mv M)$ represent its trace, determinant, and rank, respectively, while $\mv M\succeq \mv 0$ ($\mv M\preceq \mv 0$) means that $\mv M$ is positive (negative) semi-definite. $\mv I$ and $\mv 0$ denote an identity matrix and an all-zero matrix with proper dimensions, respectively. For a vector $\mv a$, $\Vert\mv a\Vert$ represents its Euclidean norm, $\mv a^{T}$ denotes its transpose, and $\text{diag}(\mv a)$ denotes a diagonal matrix whose diagonal elements are specified by $\mv a$. For a time-dependent function $x(t)$, $\dot {x}(t)$ denotes its first derivative with respect to time $t$. The notation $\log_{2}(\cdot)$ denotes the logarithm function with base 2, $e$ denotes the natural constant, and $\mathbb{E}(\cdot)$ denotes the statistic expectation.

\section{System Model}
As shown in Fig. \ref{cog_UAV_communication_picture}, we consider a new spectrum sharing scenario for UAV communications, where a cognitive/secondary UAV  transmitter communicates with a ground SR, in the presence of a set of $K\ge 1$ primary users that operate over the same frequency band. Let $\mathcal{K}\triangleq \{1,\ldots,K\}$ denote the set of ground PRs.  We focus on the cognitive UAV communication over a particular mission period, denoted by $\mathcal T=[0,T]$, with duration $T>0$ in second (s). In practice, the mission period $T$ is generally prescribed, which is set based on the UAV's maximum endurance and the requirements in different applications. Without loss of generality, we consider a 3D coordinate system with the SR located at the origin $(0,0,0)$ and each PR $k\in\mathcal K$ at a fixed location $(x_{k},y_{k},0)$, where $\mv w_{k}=(x_{k},y_{k})\in\mathbb{R}^{2\times 1}$ denotes the horizontal location of PR $k$. We consider offline optimization in this paper by assuming that the UAV perfectly knows the locations of the ground SR and PRs\footnote{As shown in Fig. \ref{cog_UAV_communication_picture}, as the PRs in both of our considered scenarios and the SR in the network-connected UAV communication scenario are ground BSs at fixed locations, their location information can be easily obtained by the UAV {\it a priori}.  In the D2D-enabled UAV-ground communication scenario, the SR can obtain its location via global positioning system (GPS) and then reports such information to the UAV.}, as well as the channel propagation environments (channel parameters) {\it a-priori} to facilitate the joint maneuver and power control design. This provides key insights and the performance upper bound for practical designs with partial/imperfect knowledge of location and channel information. In the following, we consider the 3D placement optimization and 3D trajectory optimization (jointly with power control) for {\it quasi-stationary} and {\it mobile} UAV scenarios, respectively.

\vspace{-10pt}\subsection{Quasi-Stationary UAV Scenario}
First, we consider the quasi-stationary UAV scenario, in which the UAV is placed at a fixed location $(x,y,z)$ (to be optimized later) over the communication period $T$. For notational convenience, let $\mv q=(x,y)$ and $z$ denote the UAV's horizontal location and altitude, respectively. Accordingly, the distances from the UAV to the SR and each PR $k\in\mathcal K$ are given by $d(\mv q,z)=\sqrt{z^{2}+\Vert \mv{q}\Vert^{2}}$ and $d_{k}(\mv q,z)=\sqrt{z^{2}+\Vert \mv{q}-\mv{w}_{k}\Vert^{2}}$, respectively.  Furthermore, let $H_{\min}>0$ and $H_{\max}>0$ denote the minimum and maximum flight altitudes of the UAV, respectively. Then we have $H_{\min}\leq z\leq H_{\max}$.

In practice, A2G wireless channels are normally dominated by the LoS links owing to the UAV's high flight altitude \cite{Qualcomm,LoS1,LoS2}. Therefore, we consider the LoS channel model with path-loss exponent $\alpha\geq 2$ for the wireless channels from the UAV to the SR and the PRs\footnote{Notice that the proposed methods can also be extended to handle other A2G channel models such as Rician fading and probabilistic LoS channel models. Please refer to Remark \ref{extendstatic} in Section $\text{\uppercase\expandafter{\romannumeral3}}$ and Remark \ref{extendmobile} in Section $\text{\uppercase\expandafter{\romannumeral4}}$ for details under the quasi-stationary and mobile UAV scenarios, respectively.}. As such, the UAV can easily obtain the CSI with them over time based on its own as well as their (fixed) locations. As a result, the channel power gains from the UAV to the SR and each PR $k\in\mathcal K$ are respectively expressed as
\begin{align}
h(\mv q,z)&=\beta_{u}d^{-\alpha}(\mv q,z)=\frac{\beta_{u}}{(z^{2}+\Vert \mv{q}\Vert^{2})^{\alpha/2}},\\
g_{k}(\mv q,z)&=\beta_{g,k}d_{k}^{-\alpha}(\mv q,z)=\frac{\beta_{g,k}}{(z^{2}+\Vert \mv{q}-\mv{w}_{k}\Vert^{2})^{\alpha/2}},
\end{align}
where $\beta_{u}$ and $\beta_{g,k}$ denote the reference channel power gains from the UAV to the SR and each PR $k\in\mathcal K$, respectively, including the transmit and receive antenna gains of communication nodes involved. In practice, the UAV may adjust its antenna's main lobe towards the SR to improve the cognitive communication rate, and the PRs (ground BSs) may adjust their main lobes {\it downwards} to better serve their respective primary transmitters (ground users) by reducing the co-channel interference from other primary transmitters. As a result, the A2G interference is generated and received via the side-lobes of the UAV's and the PRs' antennas, respectively. Therefore, we have $\beta_{g,k}\leq \beta_{0},~\forall k\in\mathcal K$, where $\beta_{0}$ denotes the maximum reference channel power gain from the UAV to the PRs when they are all equipped with the omnidirectional antennas.

Accordingly, by letting $p\geq 0$ denote the transmit power of the UAV, the maximum achievable rate from the UAV to the SR in bits/second/Hertz (bps/Hz) is given by
\begin{align}\label{staticrate1}
R\left(p,\mv q,z\right)&=\log_{2}\left(1+\frac{h(\mv q,z)p}{\sigma^{2}}\right)\nonumber\\
&=\log_{2}\left(1+\frac{\eta_{u}p}{(z^{2}+\Vert \mv{q}\Vert^{2})^{\alpha/2}}\right),
\end{align}
where $\sigma^{2}$ denotes the total power of receiver noise and terrestrial interference at the SR, and $\eta_{u}\triangleq \beta_{u}/\sigma^{2}$ denotes the reference signal-to-interference-plus-noise ratio (SINR). Let $P>0$ denote the maximum transmit power at the UAV. We thus have $0\leq p\leq P$.

Under spectrum sharing, the secondary UAV communication introduces A2G co-channel interference to the ground PRs, and the resultant interference power at each PR $k\in\mathcal K$ is given by $\tilde{Q}_{k}\left(p,\mv q,z\right)=g_{k}(\mv q,z)p=\frac{\beta_{g,k}p}{(z^{2}+\Vert \bm{q}-\bm{w}_{k}\Vert^{2})^{\alpha/2}}.$ As the UAV may not be able to know the exact receive antenna gain at each PR (due to the unknown receive antenna direction), we consider the worst-case A2G interference by replacing $\beta_{g,k}$ with $\beta_{0}$, $\forall k\in\mathcal K$. Thus, we have
\begin{align}\label{Q}
\tilde{Q}_{k}\left(p,\mv q,z\right)\leq Q_{k}\left(p,\mv q,z\right)=\frac{\beta_{0}p}{(z^{2}+\Vert \mv{q}-\mv{w}_{k}\Vert^{2})^{\alpha/2}},~\forall k\in\mathcal K.
\end{align}
In order to protect the primary communications, we apply the IT technique that is widely adopted in the CR literature (see, e.g., \cite{RZhangIT}), such that the received (worst-case) interference power $Q_{k}(p,\mv q,z)$ at each PR $k$ cannot exceed a maximum threshold, denoted by $\Gamma\geq 0$\footnote{In practice, each PR also suffers the terrestrial uplink interference from other co-channel terrestrial users. However, due to the more severe path-loss, shadowing, and small-scale fading over terrestrial channels, as well as the relatively mature interference mitigation techniques for terrestrial networks \cite{ICIC}, we assume that in this paper the terrestrial interference is much weaker than the A2G interference from the UAV. As a result, under our considered setup, each PR's rate performance is mainly limited by the A2G interference given in (\ref{Q}).}, i.e., $Q_{k}\left(p,\mv q,z\right)\leq \Gamma,~\forall k\in\mathcal K$, and thus we have $\beta_{0}p/(z^{2}+\Vert\mv q-\mv w_{k}\Vert^{2})^{\alpha/2}\leq \Gamma,~\forall k\in\mathcal K$\footnote{Notice that the IT constraint at each PR $k\in\mathcal K$ only depends on its location $\mv w_{k}$. Therefore, the UAV only needs to know the locations of PRs, but does not need to know the locations of primary transmitters (ground users).}.

In the quasi-stationary UAV scenario, our objective is to maximize the SR's achievable rate (i.e, $R(p,\mv q,z)$), by jointly optimizing the UAV's 3D location $\mv q$ and $z$, and transmit power $p$. The problem is formulated as
\begin{align}
\max\limits_{p,\bm{q},z}~&\log_{2}\left(1+\frac{\eta_{u}p}{(z^{2}+\Vert\mv{q}\Vert^{2})^{\alpha/2}}\right)\nonumber \\
\text{s.t.}\ &H_{\min}\leq z\leq H_{\max},\label{staticaltitude}\\
&0\leq p\leq P,\label{staticpower0}\\
&\frac{\beta_{0}p}{(z^{2}+\Vert\mv q-\mv w_{k}\Vert^{2})^{\alpha/2}}\leq \Gamma,~\forall k\in\mathcal K.\label{staticIT0}
\end{align}
Notice that the cognitive communication performance in this scenario is regardless of the mission duration $T$. Due to the monotonic increasing property of the $\log_{2}(\cdot)$ function, the above problem is equivalent to maximizing the SR's received SNR, i.e.,
\begin{align}
\text{(P1):}~\max\limits_{p,\bm{q},z}~&\frac{p}{(z^{2}+\Vert\mv{q}\Vert^{2})^{\alpha/2}}\nonumber \\
\text{s.t.}~&\text{(\ref{staticaltitude})--(\ref{staticIT0}),}\nonumber
\end{align}
where the constant $\eta_{u}$ is omitted at the objective function without loss of optimality. Note that problem (P1) is non-convex, as the objective function is non-concave and the constraints in (\ref{staticIT0}) are non-convex. Therefore, this problem is generally difficult to be solved optimally. We will tackle this problem in Section $\text{\uppercase\expandafter{\romannumeral3}}$.

\vspace{-10pt}\subsection{Mobile UAV Scenario}
Next, we consider the mobile UAV scenario, in which the UAV flies freely in the 3D space during the mission period $\mathcal T$, subject to pre-determined initial and final locations.  Suppose that the UAV has a time-varying 3D location $(\hat{x}(t),\hat{y}(t),\hat{z}(t))$ at time instant  $t\in\mathcal T$, where $\hat{\mv{{q}}}(t)=(\hat{x}(t),\hat{y}(t))$ denotes the horizontal UAV location, and $\hat{z}(t)$ denotes the flight altitude. Specifically, the UAV's initial and final horizontal locations are given as $\hat{\mv {q}}_{I}=(x_{I},y_{I})$ and $\hat{\mv {q}}_{F}=(x_{F},y_{F})$, and the corresponding altitudes are $\hat{z}_{I}$ and $\hat{z}_{F}$, respectively. Let $\hat{V}_{H}$, $\hat{V}_{A}$ and $\hat{V}_{D}$ denote the UAV's maximum horizontal speed, vertical ascending speed, and vertical descending speed in meters/second (m/s), respectively (e.g., $\hat{V}_{H}=\text{26 m/s}$, $\hat{V}_{A}=\text{6 m/s}$, and $\hat{V}_{D}=\text{4 m/s}$ for DJI's Inspire 2 drones\cite{dajiang}). Then we obtain the UAV's flying speed constraints as $\sqrt{\dot{\hat{x}}^{2}(t)+\dot{\hat{y}}^{2}(t)}\le \hat{V}_{H}$, $-\hat{V}_{D}\leq\dot{\hat{z}}(t)\leq \hat{V}_{A}$, $\forall t\in\mathcal T$. In this case, the minimum required duration for the UAV to fly straightly from the initial location to the final location is given by
$$T_{\min}\triangleq \begin{cases}
\max\left(\Vert\hat{\mv q}_{F}-\hat{\mv q}_{I}\Vert/\hat{V}_{H},|\hat{z}_{F}-\hat{z}_{I}|/\hat{V}_{A}\right),&\text{if}~\hat{z}_{F}\geq \hat{z}_{I},\\
\max\left(\Vert\hat{\mv q}_{F}-\hat{\mv q}_{I}\Vert/\hat{V}_{H},|\hat{z}_{F}-\hat{z}_{I}|/\hat{V}_{D}\right),&\text{if}~\hat{z}_{F}<\hat{z}_{I}.
\end{cases}$$
Therefore, we must have $T\geq T_{\min}$  in order for the UAV trajectory design to be feasible. For ease of exposition, we discretize the communication period $\mathcal T$ into $N$ time slots each with equal duration $\delta_{t}=T/N$, which is chosen to be sufficiently small such that the UAV's location can be assumed to be approximately constant within each time slot even at its maximum flying speed\footnote{However, if $\delta_{t}$ is chosen too small, the number of time slots $N$ will become excessively large, thus leading to prohibitive computational complexity. Therefore, $\delta_{t}$ or $N$ should be chosen to balance between the computational accuracy and complexity.}. Accordingly, let $\mv{q}[n]=(x[n],y[n])$ and $z[n]$ denote the UAV's horizontal location and altitude at time slot $n\in\mathcal N\triangleq \{1,\ldots,N\}$. Define $V_{H}=\hat{V}_{H}\delta_{t}$, $V_{A}=\hat{V}_{A}\delta_{t}$, and $V_{D}=\hat{V}_{D}\delta_{t}$. As a result, we have the following constraints on the UAV trajectory:
\begin{align}
&\Vert\mv{q}[n]-\mv{q}[n-1]\Vert\leq V_{H},\forall n\in\mathcal N\backslash\{1\},\nonumber\\
&-V_{D}\leq z[n]-z[n-1]\leq V_{A}, \forall n\in\mathcal N\backslash\{1\},\nonumber\\
&\mv{q}[1]=\hat{\mv{q}}_{I},~\mv{q}[N]=\hat{\mv{q}}_{F},~z[1]=\hat{z}_{I},~z[N]=\hat{z}_{F}.\nonumber
\end{align}

Furthermore, let $p[n]$ denote the transmit power of the UAV at time slot $n$, where $0\leq p[n]\leq P,~\forall n\in\mathcal N$. Assuming that the Doppler effect due to the UAV's mobility is perfectly compensated at the receiver based on existing techniques \cite{Doppler}, the achievable rate from the UAV to the SR in bps/Hz in this slot is expressed as $R(p[n],\mv q[n],z[n])$ in (\ref{staticrate1}). In addition, at each time slot $n\in\mathcal N$, the UAV's resultant (worst-case) interference power at each PR $k$ cannot exceed the IT threshold $\Gamma$, i.e.,
$$\frac{\beta_{0}p[n]}{(z^{2}[n]+\Vert\mv q[n]-\mv w_{k}\Vert^{2})^{\alpha/2}}\leq \Gamma,~\forall n\in\mathcal N, k\in\mathcal K.$$

Our objective is to maximize the SR's average achievable rate (i.e., $\frac{1}{N}\sum_{n=1}^{N}R(p[n],\mv q[n],z[n])$), by optimizing the UAV's time-varying 3D locations (or trajectory) $\{\mv q[n], z[n]\}$, and the transmit power allocation $\{p[n]\}$. Therefore, the problem of our interest is formulated as
\begin{align}
\text{(P2):}\ &\max\limits_{\left\{p[n],\bm{q}[n],z[n]\right\}}\frac{1}{N}\sum\limits_{n=1}^{N}\log_{2}\left(1+\frac{\eta_{u}p[n]}{(z^{2}[n]+\Vert\mv{q}[n]\Vert^{2})^{\alpha/2}}\right)\nonumber \\
\text{s.t.}~~~&\Vert\mv{q}[n]-\mv{q}[n-1]\Vert\leq V_{H},\forall n\in\mathcal N\backslash\{1\},\label{UAV trajectory1}\\
&-V_{D}\leq z[n]-z[n-1]\leq V_{A}, \forall n\in\mathcal N\backslash\{1\},\label{UAV trajectory2}\\
&\mv{q}[1]=\hat{\mv{q}}_{I},~\mv{q}[N]=\hat{\mv{q}}_{F},~z[1]=\hat{z}_{I},~z[N]=\hat{z}_{F},\label{UAV trajectory3}\\
&H_{\min}\leq z[n]\leq H_{\max},~\forall n\in\mathcal N,\label{altitudeconstraint}\\
&0\leq p[n]\leq P,~\forall n\in\mathcal N,\label{mobilepower}\\
&\frac{\beta_{0}p[n]}{(z^{2}[n]+\Vert\mv q[n]-\mv w_{k}\Vert^{2})^{\alpha/2}}\leq \Gamma,~\forall n\in\mathcal N, k\in\mathcal K.\label{mobileIT}
\end{align}
Here, (\ref{UAV trajectory1}) denotes the UAV's  maximum horizontal speed constraints, (\ref{UAV trajectory2}) denotes its maximum vertical ascending and descending speed constraints, (\ref{UAV trajectory3}) specifies the constraints on its initial and final locations, (\ref{altitudeconstraint}) denotes its flight altitude constraints, (\ref{mobilepower}) denotes its maximum transmit power constraint, and (\ref{mobileIT}) denotes the PRs' stringent IT constraints. Note that problem (P2) is non-convex, which is even more difficult to be solved than (P1) due to the involvement of time-varying optimization variables. We will propose an efficient algorithm to solve (P2) sub-optimally in Section $\text{\uppercase\expandafter{\romannumeral4}}$.

\section{Joint 3D Placement and Transmit Power Optimization in Quasi-Stationary UAV Scenario}
In this section, we derive the solution to the joint 3D placement and transmit power optimization problem (P1) in the quasi-stationary UAV scenario. To start with, we introduce the following variable transformation for the UAV's transmit power $p$, i.e., $p=\hat{p}^{\alpha/2}$, to change the objective function of (P1) into $(\hat{p}/(z^{2}+\Vert\bm q\Vert^{2}))^{\alpha/2}$. Due to the monotonic increasing property of the function $(\tilde{x})^{\alpha/2}$ with $\tilde{x}\geq 0$, the exponent $\alpha/2$ can be omitted without loss of optimality. Accordingly, with some manipulation, problem (P1) can be recast in a more tractable form, i.e.,
\begin{align}
\text{(P1.1):}~\max\limits_{\hat{p},\bm q,z}~&\frac{\hat{p}}{z^{2}+\Vert\mv q\Vert^{2}}\nonumber\\
\text{s.t.}~&\frac{\hat{\beta}_{0}\hat{p}}{z^{2}+\Vert\mv q-\mv w_{k}\Vert^{2}}\leq \hat{\Gamma},~\forall k\in\mathcal K,\label{staticIT}\\
&0\leq \hat{p}\leq \hat{P},\label{staticpower}\\
&\text{(\ref{staticaltitude})},\nonumber
\end{align}
where $\hat{\beta}_{0}=\beta_{0}^{2/\alpha}$, $\hat{\Gamma}=\Gamma^{2/\alpha}$, and $\hat{P}=P^{2/\alpha}$. As a result, the optimal solution to (P1) can be obtained by solving (P1.1) and then obtaining the optimal $p$ via the relation $p=\hat{p}^{\alpha/2}$. In the following, we first consider a simplified problem of (P1.1) with UAV's horizontal location being given to draw some useful insights. Next, we derive the UAV's optimal altitude for (P1.1) and then apply the SDR technique to obtain the optimal solution of $\mv q$ and $\hat{p}$ to (P1.1). Finally, we consider a special case of (P1.1) with only $K=1$ PR, for which the closed-form optimal solution is obtained to draw further insights.

\vspace{-10pt}\subsection{Simplified Problem Given UAV's Horizontal Location}
First, in order to gain design insights, we consider a simplified case when the UAV's horizontal location $\mv q$ is given {\it a-priori}. In this case, the original problem (P1.1) is simplified as
\begin{align}
\text{(P3):}~\max\limits_{\hat{p},z}~&\frac{\hat{p}}{z^{2}+\Vert\mv q\Vert^{2}}\nonumber\\
\text{s.t.}~&\frac{\hat{\beta}_{0}\hat{p}}{z^{2}+\Vert\mv q-\mv w_{k}\Vert^{2}}\leq \hat{\Gamma},~\forall k\in\mathcal K,\label{pzIT}\\
&\text{(\ref{staticaltitude}) and (\ref{staticpower})}.\nonumber
\end{align}
Let $\tilde{k}(\mv q)=\text{arg}\min\limits_{k\in\mathcal K}\Vert\mv q-\mv w_{k}\Vert$ denote the PR that is closest to the UAV in the horizontal direction. Then it is evident that the IT constraints for the $K$ PRs in (\ref{pzIT}) are satisfied as long as that for the $\tilde{k}(\mv q)$-th PR is ensured. Notice that if $\tilde{k}(\mv q)$ is not unique, i.e., the UAV has the same shortest distance with two or more PRs, then we can simply choose any one of these PRs as $\tilde{k}(\mv q)$ without loss of optimality. Accordingly, the IT constraints in (\ref{pzIT}) can be reduced to
\begin{align}\label{pzIT1}
\frac{\hat{\beta}_{0}\hat{p}}{z^{2}+\Vert\mv q-\mv w_{\tilde{k}(\bm q)}\Vert^{2}}\leq\hat{\Gamma}.
\end{align}
Then, we have the following proposition.
\begin{proposition}\label{pzlemma}
The optimal solution to (P3), denoted by $z^{*}(\mv q)$ and $\hat{p}^{*}(\mv q)$, is given as follows.
\begin{itemize}
\item {\it Case 1:} If $\Vert\mv q-\mv w_{\tilde{k}(\bm q)}\Vert< \Vert\mv q\Vert$ (i.e., the UAV is closer to PR $\tilde{k}(\mv q)$ than the SR), then the optimal altitude is $z^{*}(\mv q)=\min(H_{\max},\max(\sqrt{\frac{\hat{\beta}_{0}\hat{p}}{\hat{\Gamma}}-\Vert\mv q-\mv w_{\tilde{k}(\bm q)}\Vert^{2}},H_{\min}))$, and the optimal solution of $\hat{p}$ is $\hat{p}^{*}(\mv q)=\min(\frac{\hat{\Gamma}}{\hat{\beta}_{0}}(H_{\max}^{2}+\Vert\mv q-\mv w_{\tilde{k}(\bm q)}\Vert^{2}),\hat{p})$.
\item {\it Case 2:} If $\Vert\mv q-\mv w_{\tilde{k}(\bm q)}\Vert> \Vert\mv q\Vert$ (i.e., the UAV is closer to the SR than PR $\tilde{k}(\mv q)$), then the optimal altitude is $z^{*}(\mv q)=H_{\min}$, and the optimal solution of $\hat{p}$ is $p^{*}(\mv q)=\min(\frac{\hat{\Gamma}}{\hat{\beta}_{0}}(H_{\min}^{2}+\Vert\mv q-\mv w_{\tilde{k}(\bm q)}\Vert^{2}),\hat{p})$.
\item {\it Case 3:} If $\Vert\mv q-\mv w_{\tilde{k}(\bm q)}\Vert=\Vert\mv q\Vert$ (i.e., the UAV has the same distance with the SR and PR $\tilde{k}(\mv q)$), then the optimal altitude $z^{*}(\mv q)$ is non-unique and can be chosen as any value between $H_{\min}$ and $\min(H_{\max},\max(\sqrt{\frac{\hat{\beta}_{0}\hat{p}}{\hat{\Gamma}}-\Vert\mv q-\mv w_{\tilde{k}(\bm q)}\Vert^{2}},H_{\min}))$. In this case, the optimal solution of $\hat{p}$  is $\hat{p}^*(\mv q)=\min(\frac{\hat{\Gamma}}{\hat{\beta}_{0}}(z^*(\mv q)+\Vert\mv q-\mv w_{\tilde{k}(\bm q)}\Vert^{2}),\hat{p})$.
\end{itemize}
\end{proposition}
\begin{IEEEproof}
See Appendix \ref{pzlemmaproof}.
\end{IEEEproof}
By combining $z^{*}(\mv q)$ together with $p^{*}(\mv q)=\hat{p}^{(*)\alpha/2}(\mv q)$, the optimal solution to (P1) in the simplified case with the given UAV's horizontal location is finally obtained.

\vspace{-10pt}\subsection{Proposed Solution to (P1.1)}
Next, we consider the original problem (P1.1), for which we use $\mv q^{\star}$, $z^{\star}$, and $\hat{p}^{\star}$ to denote the optimal solution. We first present the following lemma.
\begin{lemma}\label{probabilityp}
At the optimal solution to (P1.1), the UAV must be placed no closer to any of the PRs than the SR, i.e., $\Vert\mv q^{\star}-\mv w_{k}\Vert\geq \Vert\mv q^{\star}\Vert,~\forall k\in\mathcal K$\footnote{Lemma \ref{probabilityp} essentially reduces the set that contains the optimal horizontal location $\mv q^{\star}$ from $\mathbb{R}^{2\times 1}$ to a smaller convex set, which is the intersection of $K$ half-spaces each specified by the inequality  $\Vert\mv q^{\star}-\mv w_{k}\Vert\geq \Vert\mv q^{\star}\Vert$ for PR $k$. However, how to search $\mv q^{\star}$ in this convex set is still challenging as shown next.}.
\end{lemma}
\begin{IEEEproof}
See Appendix \ref{probability}.
\end{IEEEproof}

Then, we have the following proposition for the optimal UAV's altitude.
\begin{proposition}\label{hmin}
$z^{\star}=H_{\min}$ is optimal for problem (P1.1), i.e., it is optimal to place the UAV at its lowest altitude.
\end{proposition}
\begin{IEEEproof}
Given UAV's horizontal location $\mv q$ as $\mv q^{\star}$, the optimal altitude solution $z^{*}(\mv q^{\star})$ to problem (P3) is identical to $z^{\star}$ to problem (P1.1). It follows from Lemma \ref{probabilityp} that $\Vert\mv q^{\star}\Vert\leq \Vert\mv q^{\star}-\mv w_{\tilde{k}({\bm q}^{\star})}\Vert$ must hold. By using this fact together with Cases 2 and 3 in Proposition \ref{pzlemma}, we have $z^{*}(\mv q^{\star})=H_{\min}$. As a result, it follows that $z^{\star}=z^{*}(\mv q^{\star})=H_{\min}$. This proposition is thus proved.
\end{IEEEproof}
\begin{remark}\label{compare}
It is interesting to compare Proposition \ref{hmin} versus Proposition \ref{pzlemma}. It is observed from Proposition \ref{hmin} that when the UAV's horizontal location is at the optimal point, the UAV should accordingly stay at its lowest altitude. This is in a sharp contrast to Proposition \ref{pzlemma}, which shows that if the UAV's horizontal location is fixed at any given point, then it is generally necessary for the UAV to adaptively adjust its altitude depending on its horizontal distances with the SR and the PRs, in order to maximize the SR's achievable rate subject to the PRs' IT constraints. In particular, when the UAV's horizontal location is closer to any PR than the SR, Proposition \ref{pzlemma} shows that the UAV may need to ascend to a higher altitude to achieve the best cognitive communication rate. This implies that if the UAV has to fly over an area with distributed PRs (e.g., for certain long-range tasks), then adjusting its flight altitude (together with horizontal location) becomes crucial for achieving the maximum cognitive UAV communication rate, especially when the UAV has to visit certain locations closer to some PRs than the SR during the flight. This result will be exploited for the 3D trajectory design in the mobile UAV scenario later.
\end{remark}

Next, it remains for us to find the optimal solution of $\hat{p}$ and $\mv q$ to problem (P1.1). By substituting $z=z^{\star}=H_{\min}$ and introducing an auxiliary value $\tau$, problem (P1.1) is re-expressed as
\begin{align}
\text{(P4):}\ \max\limits_{\tau,\hat{p},\bm{q}}~&\tau\nonumber\\
\text{s.t.}~&\Vert\mv q\Vert^{2}\leq \frac{\hat{p}}{\tau}-H_{\min}^{2},\label{p21t}\\
&\Vert\mv q-\mv w_{k}\Vert^{2}\geq \frac{\hat{\beta}_{0}\hat{p}}{\hat{\Gamma}}-H_{\min}^{2},~\forall k\in\mathcal K,\label{p21IT}\\
&0\leq \hat{p}\leq \hat{p}.\label{p21p}
\end{align}
However, problem (P4) is still non-convex\footnote{Notice that given $\mv q$, (P4) is a linear programming (LP) over $\tau$ and $\hat{p}$, thus can be optimally solved; nevertheless, the key challenge here is to jointly optimize all variables, which renders (P4) a non-convex problem.}, as constraint (\ref{p21t}) is non-convex due to the coupling between $\hat{p}$ and $\tau$, and the constraints in (\ref{p21IT}) are non-convex quadratic constraints.

First, we deal with the non-convex quadratic constraints in (\ref{p21IT}) by using  the SDR technique. Towards this end, we first equivalently recast (P4) as the following problem (P4.1) with homogeneous quadratic terms by introducing an auxiliary variable $\theta$.
\begin{align}
\text{(P4.1):}\ \max\limits_{\tau,\hat{p},\bm{q},\theta}~&\tau\nonumber\\
\text{s.t.}~&(\mv q,\theta)^{T}\mv A(\mv q,\theta)\leq \frac{\hat{p}}{\tau}-H_{\min}^{2},\label{p22t}\\
&(\mv q,\theta)^{T}\mv B_{k}(\mv q,\theta)\geq \frac{\hat{\beta}_{0}\hat{p}}{\hat{\Gamma}}-H_{\min}^{2},~\forall k\in\mathcal K,\label{p22it}\\
&\theta^{2}=1,\label{p22m}\\
&\text{(\ref{p21p}),}\nonumber
\end{align}
where $\mv A\triangleq \text{diag}((1,1,0))\in\mathbb{R}^{3\times3}$ and $\small\mv B_{k}\triangleq\begin{bmatrix}\mv I&-\mv w_{k}^{T}\\-\mv w_{k}&\Vert\mv w_{k}\Vert^{2}\end{bmatrix}\in\mathbb{R}^{3\times3},~\forall k\in\mathcal K.$

Then, by introducing $ \mv s=(\mv q,\theta)\in \mathbb{R}^{3\times 1}$ and $\mv S=\mv s\mv s^{T}\in\mathbb{R}^{3\times 3}$, with $\mv S\succeq \mv 0$ and $\text{rank}(\mv S)\le 1$, problem (P4.1) is further reformulated as
\begin{align}
\text{(P4.2):}\ \max\limits_{\tau,\hat{p},\bm S}~&\tau\nonumber\\
\text{s.t.}~&\text{Tr}(\mv A\mv S)\leq \frac{\hat{p}}{\tau}-H_{\min}^{2},\label{p23t}\\
&\text{Tr}(\mv B_{k}\mv S)\geq\frac{\hat{\beta}_{0}\hat{p}}{\hat{\Gamma}}-H_{\min}^{2},~\forall k\in\mathcal K,\label{p23it}\\
&\text{Tr}(\mv C\mv S)=1,\label{p23m}\\
&\text{rank}(\mv S)\le 1,\label{p23rank}\\
&\mv S\succeq \mv 0,\label{p23s0}\\
&\text{(\ref{p21p}),}\nonumber
\end{align}
where $\mv C\triangleq\text{diag}((0, 0,1))$. Notice that the rank constraint in (\ref{p23rank}) is non-convex. To address this issue, we relax problem (P4.2) by dropping this rank constraint, and denote the relaxed problem of (P4.2) as (P4.3).

Next, we consider problem (P4.3).  Although constraint (\ref{p23t}) is non-convex due to the coupling between $\hat{p}$ and $\tau$, problem (P4.3) can be solved by equivalently solving the following feasibility problems (P4.4.$\tau$) under any given $\tau\geq 0$, together with a bisection search over $\tau$.
\begin{align}
\text{(P4.4.$\tau$):}~\text{Find}~&\hat{p}, \mv S\nonumber\\
\text{s.t.}~&\text{(\ref{p21p}), (\ref{p23t}), (\ref{p23it}), (\ref{p23m}), and (\ref{p23s0}).}\nonumber
\end{align}
In particular, denote by $\tau^{\star}$ the optimal solution of $\tau$ to (P4.3). Then, under any given $\tau\geq 0$, we have $\tau\leq \tau^{\star}$ if problem (P4.4.$\tau$) is feasible; otherwise, we have $\tau>\tau^{\star}$. Therefore, we can solve (P4.3) by using the bisection search\footnote{Suppose that the searching range of $\tau$ is an interval $[0, \tau_{\max}]$. As such, the maximum number of iterations for bisection search is given by $\lceil\log_{2}(\tau_{\max}/\epsilon)\rceil$, where $\epsilon$ is a positive constant that controls the accuracy, and $\lceil\tilde{y}\rceil$ denotes the minimum integer that is no smaller than $\tilde{y}$. Since the number of required iterations is a logarithmic function with respect to $\tau_{\max}/\epsilon$, the convergence of the bisection search is exponentially fast.} over $\tau\geq 0$, and checking the feasibility of problem (P4.4.$\tau$) under any given $\tau\geq 0$ \cite{Convex}. Notice that under any given $\tau\geq 0$, problem (P4.4.$\tau$) is a semi-definite program (SDP) that is convex, and thus can be optimally solved by using standard convex optimization techniques,  such as the interior point method\cite{Convex}. With the obtained $\tau^{\star}$, suppose that the corresponding feasible/optimal solution to problem (P4.4.$\tau$) is $\hat{p}^{\star}(\tau)$ and $\mv S^{\star}(\tau)$. Accordingly, they are also the optimal solution to problem (P4.3), denoted by $\hat{p}^{\star}$ and $\mv S^{\star}$.

Now, it still remains to construct the optimal solution to (P4.2), or equivalently (P4.1) and (P4). In particular, if $\text{rank}(\mv S^{\star})\leq 1$, then the SDR is tight. In this case, the solution of $\hat{p}^{\star}$ and $\mv S^{\star}$ are also the optimal solution to (P4.2). By performing the eigenvalue decomposition (EVD) for $\mv S^{\star}$, we can obtain the optimal solution $\mv s^{\star}=(\mv q^{\star},\theta^{\star})$ to (P4.1) and (P4) as the dominant eigenvector of $\mv S^{\star}$. Accordingly, $\hat{p}^{\star}$ is also the optimal solution of $\hat{p}$ to (P4.1) and (P4). However, if $\text{rank}(\mv S^{\star})>1$, then we need to construct a rank-one solution of $\mv S$ to (P4.2) via additional processing such as the Gaussian randomization procedure that is widely adopted in the SDR literature (see, e.g., \cite{SDR}). Fortunately, in our simulations with randomly generated PRs' locations, the optimal solution of $\mv S^{\star}$ to problem (P4.3) is always rank-one. Therefore, the Gaussian randomization procedure is not required in general. More specifically, we can rigorously prove that the optimal solution of $\mv S^{\star}$ to (P4.3) is rank-one in the following special case, though our proposed SDR-based solution is applicable for the general case with any PRs' locations.
\begin{proposition}\label{rankone}
When the $K$ PRs are located at the same side of the SR\footnote{There are in total four cases when the $K$ PRs are located at the same side of the SR: 1) $x_{k}\geq 0$, $\forall k\in\mathcal K$, and there exists at least one PR $\bar{k}\in\mathcal K$ with $x_{\bar k}>0$; 2) $x_{k}\leq 0$, $\forall k\in\mathcal K$, and there exists at least one PR $\bar{k}\in\mathcal K$ with $x_{\bar k}<0$; 3) $y_{k}\geq 0$, $\forall k\in\mathcal K$ , and there exists at least one PR $\bar{k}\in\mathcal K$ with $y_{\bar k}>0$; and 4) $y_{k}\leq 0$, $\forall k\in\mathcal K$, and there exists at least one PR $\bar{k}\in\mathcal K$ with $y_{\bar k}<0$.}, it follows that the optimal solution $\mv S^{\star}$ to (P4.3) is always rank-one.
\end{proposition}
\begin{IEEEproof}
See Appendix \ref{rankoneproof}.
\end{IEEEproof}
Therefore, the solution to (P4) is finally obtained as $\hat{p}^{\star}$ and $\mv q^{\star}$. By combining them together with $z^{\star}=H_{\min}$, problem (P1.1) is solved. As a result, the optimal solution to (P1) is finally obtained as $z^{\star}$, $\mv q^{\star}$, and $p^{\star}=\hat{p}^{(\star)\alpha/2}$.
\vspace{-10pt}\subsection{Special Case with $K=1$ PR}
In this subsection, we consider the special case of (P1.1) with $K=1$ PR and derive the closed-form optimal solution to gain additional insights. In this case, by substituting $z=z^{\star}=H_{\min}$, problem (P1.1) is simplified as
\begin{align}
\text{(P5):}~\max\limits_{\hat{p},\bm{q}}~&\frac{\hat{p}}{H_{\min}^{2}+\Vert\mv q\Vert^{2}}\nonumber\\
\text{s.t.}~&\frac{\hat{\beta}_{0}\hat{p}}{H_{\min}^{2}+\Vert\mv{q}-\mv w_{1}\Vert^{2}}\leq \hat{\Gamma},\label{IT1PR}\\
&\text{(\ref{staticpower})}.\nonumber
\end{align}
For (P5), we first have the following lemma.

\begin{lemma}\label{structure}
Under any feasible $\hat{p}$, the optimal solution of $\mv q$ to problem (P5) is given by $\mv q=-a \frac{\mv w_{1}}{\Vert\mv w_{1}\Vert}$, where $a\geq 0$ denotes the horizontal distance between the UAV and the SR.
\end{lemma}

\begin{IEEEproof}
See Appendix \ref{structureproof}.
\end{IEEEproof}
From Lemma \ref{structure}, it is evident that at the optimality, the UAV should be placed above a point at the PR's opposite direction along the line connecting the SR and the PR, to minimize the interference to the PR. By substituting $\mv q=-a\frac{\mv w_{1}}{\Vert\mv w_{1}\Vert}$, problem (P5) is re-expressed as
\begin{align}
\text{(P5.1):}~\max\limits_{\hat{p},a\geq 0}~&\frac{\hat{p}}{H_{\min}^{2}+a^{2}}\nonumber\\
\text{s.t.}~&\frac{\hat{\beta}_{0}\hat{p}}{H_{\min}^{2}+(a+\Vert\mv w_{1}\Vert)^{2}}\leq \hat{\Gamma},\label{IT1PR1}\\
&\text{(\ref{staticpower})}.\nonumber
\end{align}

For convenience, we define $p_{1}\triangleq \frac{\hat{\Gamma}}{\hat{\beta}_{0}}(\Vert\mv w_{1}\Vert^{2}+H_{\min}^{2})$, which denotes the UAV's maximally allowable value of $\hat{p}$ for the IT constraint (\ref{IT1PR}) to be feasible, when the UAV is located exactly above the SR at $(0,0,H_{\min})$. Then, we have the following proposition.
\begin{proposition}\label{1PRhover2}
The optimal solution of $\hat{p}$ to problem (P5.1) is given by $\hat{p}^{\star}=\min(\hat{p},\tilde{p}^{\star})$, where
\vspace{-5pt}\begin{align}
\tilde{p}^{\star}\triangleq \frac{\hat{\Gamma}}{\hat{\beta}_{0}}\left(\frac{\left(\Vert\mv w_{1}\Vert+\sqrt{\Vert\mv w_{1}\Vert^{2}+4H_{\min}^{2}}\right)^{2}}{4}+H_{\min}^{2}\right).\label{optimalp1}
\end{align}
Accordingly, the optimal solution of $a$ to problem (P5.1) is given as
\begin{equation}\label{optimala1}
a^{\star}=\begin{cases}
\tilde{a}^{\star}\triangleq\frac{\sqrt{\Vert\mv w_{1}\Vert^{2}+4H_{\min}^{2}}-\Vert\mv w_{1}\Vert}{2},&\hat{p}> \tilde{p}^{\star},\\
\sqrt{\frac{\hat{\beta}_{0}P}{\hat{\Gamma}}-H_{\min}^{2}}-\Vert\mv w_{1}\Vert<\tilde{a}^{\star},&p_{1}\leq\hat{p}\leq \tilde{p}^{\star},\\
0,&\hat{p}<p_{1},
\end{cases}
\end{equation}
\end{proposition}
\begin{IEEEproof}
See Appendix \ref{1PRhover2proof}.
\end{IEEEproof}
By combining Proposition \ref{1PRhover2} and Lemma \ref{structure}, the optimal solution to problem (P5) is finally obtained as $\hat{p}^{\star}$ and $\mv q^{\star}=-a^{\star}\frac{\mv w_{1}}{\Vert\mv w_{1}\Vert}$. As a result, the optimal solution to (P1) in the special case with $K=1$ PR is finally obtained as $z^{\star}$, $\mv q^{\star}$, and $p^{\star}=\hat{p}^{(\star)\alpha/2}$.

Proposition \ref{1PRhover2} provides interesting insights on the optimal horizontal location and transmit power solution to the special case of (P1) with $K=1$ PR. Firstly, when the UAV's maximum transmit power $P$ is sufficiently large, the UAV should transmit at an optimized power $p^{\star}=\tilde{p}^{(\star)\alpha/2}$ (with $\tilde{p}^{\star}$ given in (\ref{optimalp1})) and be placed at an optimized horizontal location with distance $\tilde{a}^{\star}$ given in (\ref{optimala1}) from the SR. Notice that $\tilde{a}^{\star}$ is only dependent on $\Vert\mv w_{1}\Vert$ and $H_{\min}$ but irrelevant to $P$; as $H_{\min}$ increases and/or $\Vert\mv w_{1}\Vert$ decreases, $\tilde{a}^{\star}$ becomes larger and thus the UAV needs to be placed further away from the SR for maximizing the SR's achievable rate subject to the IT constraint. Furthermore, when $P$ becomes smaller with $\hat{p}\leq \tilde{p}^{(\star)}$, the UAV should transmit at the full power $P$ and be placed at a horizontal location closer to the SR. In addition, if $P$ becomes sufficiently small with $\hat{p}<p_{1}$, then the UAV should be placed exactly above the SR and transmit with full power $P$.

\begin{remark}\label{extendstatic}
Notice that although in this paper we consider the LoS channel model, the design principles are applicable to other stochastic A2G channel models such as Rician fading and probabilistic LoS channels.

Specifically, denote by $\tilde{h}(\mv q,z)$ and $\tilde{g}_{k}(\mv q,z)$ the instantaneous channel power gains from the UAV to the SR and to the PR $k\in\mathcal K$, respectively, which are random variables whose probability density functions generally depend on the elevation angles between the UAV and the ground nodes. In general, as the UAV altitude $z$ increases, the $K$-factor becomes larger for Rician fading channel\cite{LoS2} and the LoS probability increases for probabilistic LoS channel\cite{plos}. For convenience, we denote $\hat{h}(\mv q,z)=\mathbb{E}(\tilde{h}(\mv q,z))$ and $\hat{g}_{k}(\mv q,z)=\mathbb{E}(\tilde{g}_{k}(\mv q,z))$ as their mean values. Then, we can maximize the average rate from the UAV to the SR, i.e., $\mathbb{E}\left(\log_{2}\left(1+\frac{p \tilde{h}(\bm q,z)}{\sigma^{2}}\right)\right)$, subject to the average IT constraints at all PRs, i.e., $\mathbb{E}\left(p\tilde{g}_{k}(\mv q,z)\right)= p\hat{g}_{k}(\mv q,z)\leq \Gamma, \forall k\in\mathcal K.$
In general, we can adopt the exhaustive search over the 3D space to find the optimal solution to this new problem.

However, due to the concavity of the $\log(\cdot)$ function, it follows from the Jensen's inequality \cite{plos} that
\begin{align}
\mathbb{E}\left(\log_{2}\left(1+\frac{p \tilde{h}(\mv q,z)}{\sigma^{2}}\right)\right)&\leq \log_{2}\left(1+\frac{p\mathbb{E}(\tilde{h}(\mv q,z))}{\sigma^{2}}\right)\nonumber\\
&=\log_{2}\left(1+\frac{p\hat{h}(\mv q,z)}{\sigma^{2}}\right).\nonumber
\end{align}
Since the channel power gains achieve their maximum under the LoS channel model, we have $\hat{h}(\mv q,z)\leq h(\mv q,z)$ and $\hat{g}_{k}(\mv q,z)\leq g_{k}(\mv q,z),~\forall k\in\mathcal K$, which lead to $\log_{2}\left(1+\frac{p\hat{h}(\bm q,z)}{\sigma^{2}}\right)\leq \log_{2}\left(1+\frac{ph(\bm q,z)}{\sigma^{2}}\right)$, i.e., the achievable rate under the LoS channel serves as an upper bound for the average achievable rate at the SR under the stochastic channel models. Similarly, the average interference power at each PR satisfies  $p\hat{g}_{k}(\mv q,z)\leq pg_{k}(\mv q,z),~\forall k\in\mathcal K.$ As a result, the optimal solution to our considered problem (P1) can be viewed as an approximate solution to the problem in the stochastic channel models. In particular, such approximations become more accurate when the $K$-factor is larger for Rician fading channel or the LoS probability is higher for probabilistic LoS channel.
\end{remark}

\section{Joint 3D Trajectory and Transmit Power Optimization in Mobile UAV Scenario}
In this section, we consider the joint UAV 3D trajectory and transmit power optimization problem (P2) in the mobile UAV scenario. To tackle this problem, we use the SCA technique to obtain a locally optimal solution.

To facilitate the implementation of SCA, we first obtain the optimal transmit power levels under any given feasible UAV trajectory $\{\mv q[n],z[n]\}$, for which the problem is expressed as
\begin{align}
\text{(P6):}\ &\max\limits_{\{p[n]\}}~\frac{1}{N}\sum\limits_{n=1}^{N}\log_{2}\left(1+\frac{\eta_{u}p[n]}{(z^{2}[n]+\Vert\mv{q}[n]\Vert^{2})^{\alpha/2}}\right)\nonumber\\
\text{s.t.}\ &0\leq p[n]\leq P,~\forall n\in\mathcal N,\label{p5power}\\
&\frac{\beta_{0}p[n]}{(z^{2}[n]+\Vert\mv q[n]-\mv w_{k}\Vert^{2})^{\alpha/2}}\leq \Gamma,~\forall n\in\mathcal N,~k\in\mathcal K.\label{p5it}
\end{align}
It is easy to show that problem (P6) can be decomposed into the following $N$ subproblems each for one slot $n\in\mathcal N$, for which the coefficient $1/N$ is ignored for brevity.
\begin{align}
\text{(P6.1.$n$):}\ &\max\limits_{p[n]\geq 0}~\log_{2}\left(1+\frac{\eta_{u}p[n]}{(z^{2}[n]+\Vert\mv{q}[n]\Vert^{2})^{\alpha/2}}\right)\nonumber\\
\text{s.t.}\ &p[n]\leq \min(P,\min\limits_{k\in\mathcal K}\frac{\Gamma}{\beta_{0}}\left(z^{2}[n]+\Vert\mv q[n]-\mv w_{k}\Vert^{2}\right)^{\alpha/2}).\label{p_power_optimization}
\end{align}
It is evident that the optimality of problem (P6.1.$n$) is attained when constraint (\ref{p_power_optimization}) is tight. Therefore, we have the optimal solution to (P6.1.$n$)'s and (P6) as

\vspace{-10pt}\begin{small}\begin{align}\label{poweroptimization}
p[n]=\min\left(P,\min\limits_{k\in\mathcal K}\frac{\Gamma}{\beta_{0}}\left(z^{2}[n]+\Vert\mv q[n]-\mv w_{k}\Vert^{2}\right)^{\alpha/2}\right),\forall n\in\mathcal N.
\end{align}\end{small}By substituting (\ref{poweroptimization}) into the objective function of (P2), problem (P2) is reformulated as
\begin{align}
\text{(P7):}\ &\max\limits_{\left\{\bm q[n],z[n]\right\}}~\frac{1}{N}\sum\limits_{n=1}^{N}\hat{R}(\mv q[n],z[n])\nonumber\\
\text{s.t.}~&\Vert\mv{q}[n]-\mv{q}[n-1]\Vert\leq V_{H},\forall n\in\mathcal N\backslash\{1\},\label{UAV trajectory1_P6}\\
&-V_{D}\delta_{t}\leq z[n]-z[n-1]\leq V_{A}, \forall n\in\mathcal N\backslash\{1\},\label{UAV trajectory2_P6}\\
&\mv{q}[1]=\hat{\mv{q}}_{I},~\mv{q}[N]=\hat{\mv{q}}_{F},~z[1]=\hat{z}_{I},~z[N]=\hat{z}_{F},\label{UAV trajectory3_P6}\\
&H_{\min}\leq z[n]\leq H_{\max},~\forall n\in\mathcal N,\label{altitudeconstraint_P6}
\end{align}
where

\vspace{-10pt}
\begin{small}
\begin{align}
&\hat{R}(\mv q[n],z[n])\nonumber\\
&=\log_{2}\left(1+\frac{\eta_{u}\min\left(P,\min\limits_{k\in\mathcal K}\frac{\Gamma}{\beta_{0}}\left(z^{2}[n]+\Vert\mv q[n]-\mv w_{k}\Vert^{2}\right)^{\alpha/2}\right)}{(z^{2}[n]+\Vert\mv q[n]\Vert^{2})^{\alpha/2}}\right).
\end{align}\end{small}Next, to solve problem (P7), we introduce two sets of auxiliary variables $\{\zeta_{1}[n]\}_{n=1}^{N}$ and $\{\zeta_{2}[n]\}_{n=1}^{N}$, and define $\tilde{R}(\zeta_{1}[n],\zeta_{2}[n])=\log_{2}(1+\eta_{u}\zeta_{1}[n]/\zeta_{2}[n])$. Accordingly, we reformulate problem (P7) as
\begin{align}
\text{(P7.1):}\ &\max\limits_{\left\{\bm q[n],z[n],\zeta_{1}[n],\zeta_{2}[n]\right\}}~\frac{1}{N}\sum\limits_{n=1}^{N}\tilde{R}(\zeta_{1}[n],\zeta_{2}[n])\nonumber\\
\text{s.t.}~~&0\leq \zeta_{1}[n]\leq P,~\forall n\in\mathcal N,\label{Pphi}\\
&\zeta_{1}[n]\leq \frac{\Gamma}{\beta_{0}}\left(z^{2}[n]+\Vert\mv q[n]-\mv w_{k}\Vert^{2}\right)^{\alpha/2},\nonumber\\
&\forall n\in\mathcal N,~k\in\mathcal K,\label{IT}\\
&(\Vert\mv q[n]\Vert^{2}+z^{2}[n])^{\alpha/2}\leq \zeta_{2}[n],~\forall n\in\mathcal N,\label{phi2}\\
&\text{(\ref{UAV trajectory1_P6})--(\ref{altitudeconstraint_P6}).}\nonumber
\end{align}
It is easy to verify that at the optimality of (P7.1), constraint $(\Vert\mv q[n]\Vert^{2}+z^{2}[n])^{\alpha/2}\leq \zeta_{2}[n]$ must hold with equality for any $n\in\mathcal N$, since otherwise, we can decrease $\zeta_{2}[n]$ to achieve a higher objective value of (P7.1)  without violating this constraint. Notice that problem (P7.1) is still non-convex, as the objective function is non-concave and the constraints in (\ref{IT}) are non-convex. To tackle this problem, we adopt the SCA technique to obtain a locally optimal solution to (P7.1) in an iterative manner. The key idea of SCA is that given a local point at each iteration, we approximate the non-concave objective function (or non-convex constraints) into a concave objective function (or convex constraints), in order to obtain an approximate convex optimization problem. By iteratively solving a sequence of approximate convex problems, we can obtain an efficient solution to the original non-convex optimization problem (P7.1).

Specifically, suppose that $\{\mv q^{(j)}[n],z^{(j)}[n],\zeta_{1}^{(j)}[n],\zeta_{2}^{(j)}[n]\}$ corresponds to the local point at the $j$-th iteration with $j\geq 1$, where $\{\mv q^{(0)}[n],z^{(0)}[n],\zeta_{1}^{(0)}[n],\zeta_{2}^{(0)}[n]\}$ corresponds to the initial point. In the following, we explain how to approximate the objective function of (P7.1) and the constraints in (\ref{IT}), respectively. First, we rewrite the objective function of (P7.1) as
\begin{align}\label{Rphi1}
\tilde{R}(\zeta_{1}[n],\zeta_{2}[n])=\log_{2}\left(\zeta_{2}[n]+\eta_{u}\zeta_{1}[n]\right)-\log_{2}(\zeta_{2}[n]).
\end{align}
Note that the objective function in (\ref{Rphi1}) is still non-concave, as $-\log_{2}\left(\zeta_{2}[n]\right)$ is non-concave. However, $-\log_{2}\left(\zeta_{2}[n]\right)$ is convex with respect to $\{\zeta_{2}[n]\}$. Notice that any convex function is globally lower-bounded by its first-order Taylor expansion at any point \cite{Convex}. Therefore, with given local point $\{\zeta_{2}^{(j)}[n]\}$ in the $j$-th iteration, $j\ge 0$, it follows that $\tilde{R}\left(\zeta_{1}[n],\zeta_{2}[n]\right)\geq \tilde{R}^{\text{lb}}\left(\zeta_{1}[n],\zeta_{2}[n]\right)$, where
\begin{align}
\tilde{R}^{\text{lb}}(\zeta_{1}[n],\zeta_{2}[n])&\triangleq\log_{2}\left(\zeta_{2}[n]+\eta_{u}\zeta_{1}[n]\right)-\log_{2}(\zeta_{2}^{(j)}[n])\nonumber\\
&-\frac{(\zeta_{2}[n]-\zeta_{2}^{(j)}[n])\log_{2}(e)}{\zeta_{2}^{(j)}[n]}.\label{Rlb}
\end{align}

Next, we consider the non-convex constraints in (\ref{IT}). Since $(\Vert\mv{q}[n]-\mv{w}_{k}\Vert^{2}+z^{2}[n])^{\alpha/2}$ is a convex function with respect to $\{\mv{q}[n],z[n]\}$, we have the following inequalities by applying the first-order Taylor expansion at any given point $\{\mv{q}^{(j)}[n],z^{(j)}[n]\}$:
\begin{align}
&(\Vert\mv{q}[n]-\mv{w}_{k}\Vert^{2}+z^{2}[n])^{\alpha/2}\geq\left(\Vert\mv{q}^{(j)}[n]-\mv{w}_{k}\Vert^{2}+z^{(j)2}[n]\right)^{\alpha/2}\nonumber\\
&+\alpha\left(\Vert\mv q^{(j)}[n]-\mv w_{k}\Vert^{2}+z^{(j)2}[n]\right)^{\frac{\alpha}{2}-1}\nonumber\\
&\left((\mv q^{(j)}[n]-\mv w_{k})^{T}(\mv q[n]-\mv q^{(j)}[n])+z^{(j)}[n](z[n]-z^{(j)}[n])\right),\nonumber\\
&\forall n\in\mathcal N,~k\in\mathcal K.
\label{SCA2}\end{align}By replacing $(\Vert \mv q[n]-\mv w_{k}\Vert^{2}+z^{2}[n])^{\alpha/2}$ in (\ref{IT}) as the right-hand-side (RHS) of (\ref{SCA2}), we approximate (\ref{IT}) as the following convex constraints:
\begin{align}
&\zeta_{1}[n]\leq\frac{\Gamma}{\beta_{0}}((\Vert\mv q^{(j)}[n]-\mv w_{k}\Vert^{2}+z^{(j)2}[n])^{\alpha/2}\nonumber\\
&+\alpha(\Vert\mv q^{(j)}[n]-\mv w_{k}\Vert^{2}+z^{(j)2}[n])^{\alpha/2-1}\nonumber\\
&((\mv q^{(j)}[n]-\mv w_{k})^{T}(\mv q[n]-\mv q^{(j)}[n])+z^{(j)}[n](z[n]-z^{(j)}[n]))),\nonumber\\
&\forall n\in\mathcal N,~k\in\mathcal K.\label{ITphi}
\end{align}To summarize, by replacing $\tilde{R}\left(\zeta_{1}[n],\zeta_{2}[n]\right)$ in the objective function as $\tilde{R}^{\text{lb}}\left(\zeta_{1}[n],\zeta_{2}[n]\right)$ in (\ref{Rlb}), and replacing the constraints in (\ref{IT}) as those in (\ref{ITphi}), problem (P7.1) is approximated as the following convex optimization problem (P7.2) at any local point $\{\mv q^{(j)}[n],z^{(j)}[n],\zeta_{1}^{(j)}[n],\zeta_{2}^{(j)}[n]\}$, which can be solved via standard convex optimization techniques such as the interior point method \cite{Convex}, with the optimal solution denoted as $\{\mv q^{(j)*}[n]\}$, $\{z^{(j)*}[n]\}$, $\{\zeta_{1}^{(j)*}[n]\}$ and $\{\zeta_{2}^{(j)*}[n]\}$.
\begin{align}
(\text{P7.2}):\ &\max\limits_{\{\bm{q}[n],z[n],\zeta_{1}[n],\zeta_{2}[n]\}}~\frac{1}{N}\sum\limits_{n=1}^{N}\tilde{R}^{\text{lb}}\left(\zeta_{1}[n],\zeta_{2}[n]\right)\nonumber\\
\text{s.t.}~~~~~&\text{(\ref{UAV trajectory1_P6})},~\text{(\ref{UAV trajectory2_P6})},~\text{(\ref{UAV trajectory3_P6})},~\text{(\ref{altitudeconstraint_P6})},~\text{(\ref{Pphi})},~\text{(\ref{phi2})},~\text{and}~\text{(\ref{ITphi})}.\nonumber
\end{align}
With the convex optimization problem (P7.2) at hand, we can obtain an efficient iterative algorithm to solve (P7.1), explained as follows. In the $j$-th iteration, the algorithm solves the convex optimization problem (P7.2) at the local point $\{\mv q^{(j)}[n],z^{(j)}[n],\zeta_{1}^{(j)}[n],\zeta_{2}^{(j)}[n]\}$, where
$\{\mv q^{(j)}[n],z^{(j)}[n]$, $\zeta_{1}^{(j)}[n], \zeta_{2}^{(j)}[n]\}$ corresponds to the optimal solution to (P7.2) obtained in the $(j-1)$-th iteration, i.e., $\mv q^{(j)}[n]=\mv q^{(j-1)*}[n]$, $z^{(j)}[n]=z^{(j-1)*}[n]$, $\zeta_{1}^{(j)}[n]=\zeta_{1}^{(j-1)*}[n]$, and $\zeta_{2}^{(j)}[n]=\zeta_{2}^{(j-1)*}[n]$, $\forall n\in\mathcal N$. We summarize this algorithm in Table \ref{SCA} as  Algorithm 1. Denote the obtained solution to (P7) as $\{\mv q^{*}[n],z^{*}[n]\}$. By substituting $\mv q^{*}[n]$ and $z^{*}[n]$ into (\ref{poweroptimization}), the corresponding transmit power is $p^{*}[n]=\min(P,\min\limits_{k\in\mathcal K}\frac{\Gamma}{\beta_{0}}\left(z^{*2}[n]+\Vert\mv q^{*}[n]-\mv w_{k}\Vert^{2}\right)^{\alpha/2})$, $\forall n\in\mathcal N$. By combining $\{p^{*}[n]\}$, $\{\mv q^{*}[n]\}$, and $\{z^{*}[n]\}$, the solution to (P2) by SCA is finally obtained.

\begin{table}\scriptsize
\caption{Algorithm 1 for Solving Problem (P7.1)} \centering
\begin{tabular}{|p{8cm}|}
\hline
\begin{itemize}
\item[a)]{\bf Initialization:} Set the initial UAV trajectory as $\{\mv q^{(0)}[n],z^{(0)}[n]\}_{n=1}^{N}$, $\zeta_{2}^{(0)}[n]=(\Vert\mv q^{(0)}[n]\Vert^{2}+z^{(0)2}[n])^{\alpha/2},~\forall n\in\mathcal N$, and $j=0$.
\item[b)] {\bf Repeat:}
\begin{itemize}
\item[1)] Solve problem (P7.2) to obtain the optimal solution as $\{\mv q^{(j)*}[n]\}_{n=1}^{N}$, $\{z^{(j)*}[n]\}_{n=1}^{N}$, $\{\zeta_{1}^{(j)*}[n]\}_{n=1}^{N}$, and $\{\zeta_{2}^{(j)*}[n]\}_{n=1}^{N}$.
\item[2)] Update the trajectory as $\mv q^{(j+1)}[n]=\mv q^{(j)*}[n]$ and $z^{(j+1)}[n]=z^{(j)*}[n]$, $\zeta_{1}^{(j+1)}[n]=\zeta_{1}^{(j)*}[n]$, and $\zeta_{2}^{(j+1)}[n]=\zeta_{2}^{(j)*}[n],~\forall n\in\mathcal N$.
\item[3)] Update $j=j+1$.
\end{itemize}
\item[c)] {\bf Until} the objective value of (P7.2) converges within a given accuracy or a maximum number of iterations is reached.
\end{itemize}\\ \hline
\end{tabular}\label{SCA}
\vspace{-10pt}\end{table}

Similarly as in \cite{RZhangrelay}, it can be shown that in Algorithm 1, after each iteration $j$, the objective function of (P7.2) achieved by $\{\mv q^{(j)}[n],z^{(j)}[n],\zeta_{1}^{(j)}[n],\zeta_{2}^{(j)}[n]\}$ is monotonically non-decreasing. As the optimal value of problem (P7.1) is upper-bounded by a finite value, it is evident that Algorithm 1 can converge to a locally optimal solution to problem (P7.1) (and thus (P2)).

Denote by $D$ the total number of iterations required in Algorithm 1. In each iteration, the convex optimization problem (P7.2) is solved via the standard interior-point method with the complexity of $O(N^{3.5}K^{1.5})$ \cite{complexity}. As a result, the overall complexity of Algorithm 1 is $O(DN^{3.5}K^{1.5})$, which is polynomial.  Also note that we consider the offline optimization for the joint trajectory and power design. Thus, Algorithm 1 only needs to be implemented in an offline manner prior to the UAV flight.
\begin{remark}\label{initial}
In order to efficiently implement Algorithm 1 for solving (P7.2) as well as (P2), we need to properly design an initial UAV trajectory. Notice that the optimal UAV location $(\mv q^{\star},z^{\star})$ to (P1) obtained in Section $\text{\uppercase\expandafter{\romannumeral3}}$ is quite efficient for maximizing the SR's communication rate while controlling the interference at the PRs. Therefore, we intuitively design a {\it fly-hover-fly} (FHF) trajectory, where the UAV first flies straightly from the initial location $(\hat{\mv q}_{I},\hat{z}_{I})$ to the optimal UAV location $(\mv q^{\star},z^{\star})$, then hovers at this point for a certain time duration, and finally flies straightly towards the final location $(\hat{\mv q}_{F},\hat{z}_{F})$. To prolong the hovering duration for improving the SR's performance, the UAV should fly at the maximum horizontal speed $\hat{V}_{H}$ and maximum ascending/descending speed $\hat{V}_{D}$/$\hat{V}_{A}$ during the flight (notice that we have $z^{\star}=H_{\min}$). As a result, we obtain the flight duration as
\begin{align}
T_{\text{fly}}&=\max(|\hat{z}_{I}-z^{\star}|/\hat{V}_{D},\Vert\mv q^{\star}-\hat{\mv q}_{I}\Vert/\hat{V}_{H})\nonumber\\
&+\max(|\hat{z}_{F}-z^{\star}|/\hat{V}_{A},\Vert\hat{\mv q}_{F}-\mv q^{\star}\Vert/\hat{V}_{H}),\nonumber
\end{align} and the hovering duration as $T-T_{\text{fly}}$. Notice that the proposed initial trajectory is only applicable when the mission duration $T$ is no smaller than $T_{\text{fly}}$. When $T_{\min}\leq T<T_{\text{fly}}$, we instead use the straight flight as the initial UAV trajectory, in which the UAV flies directly from the initial location to the final location at a constant horizontal speed $\tilde{V}_{H}=\Vert\hat{\mv q}_{F}-\hat{\mv q}_{I}\Vert/T$ and a constant vertical speed $\tilde{V}_{L}=|\hat{z}_{F}-\hat{z}_{I}|/T$.
\end{remark}

\begin{remark}\label{extendmobile}
The design principles used in this section are also applicable to other stochastic channel models such as Rician fading and probabilistic LoS channels. Similarly as in the quasi-stationary UAV scenario, we consider the average rate performance of the considered system. Specifically, in the objective function of (P2), the achievable rate under the deterministic LoS channel for each slot $n$ can be replaced with the SR's average rate over the same slot. Additionally, in the IT constraints in (\ref{mobileIT}), the PR's received interference power at each slot $n$ is modified as the average interference power over the same slot. Our proposed solution under the LoS channel then provides an efficient approximate solution to this new problem, while such approximations become more accurate when the $K$-factor is larger for Rician fading channel or the LoS probability is higher for probabilistic LoS channel. Alternatively, we can also introduce a homogenous approximation to the $K$-factor or LoS probability by assuming that they are constant throughout the UAV's flight in the stochastic channel models (see e.g., \cite{plos}). Accordingly, the resulting problem has the same form as (P2), for which we can adopt a similar SCA-based algorithm to obtain a converged solution.
\end{remark}

\section{Numerical Results}
In this section, we present numerical results to validate the performance of our proposed  joint design of UAV's  maneuver and transmit power. Unless otherwise stated, we set the noise power at the SR (including the background interference and noise) as $\sigma^{2}=\text{--80 dBm}$, the reference channel power gain at the SR as $\beta_{u}=\text{--30 dB}$, the maximum reference channel power gain from the UAV to PRs as $\beta_{0}=\text{--30 dB}$, the path-loss exponent as $\alpha=2$, and the UAV's minimum and maximum flight altitudes as $H_{\min}=\text{170 m}$  and $H_{\max}=\text{220 m}$ \cite{LoS1}, respectively.
\begin{figure*}
\begin{minipage}[t]{0.48\linewidth}
\centering
\includegraphics[width=7cm]{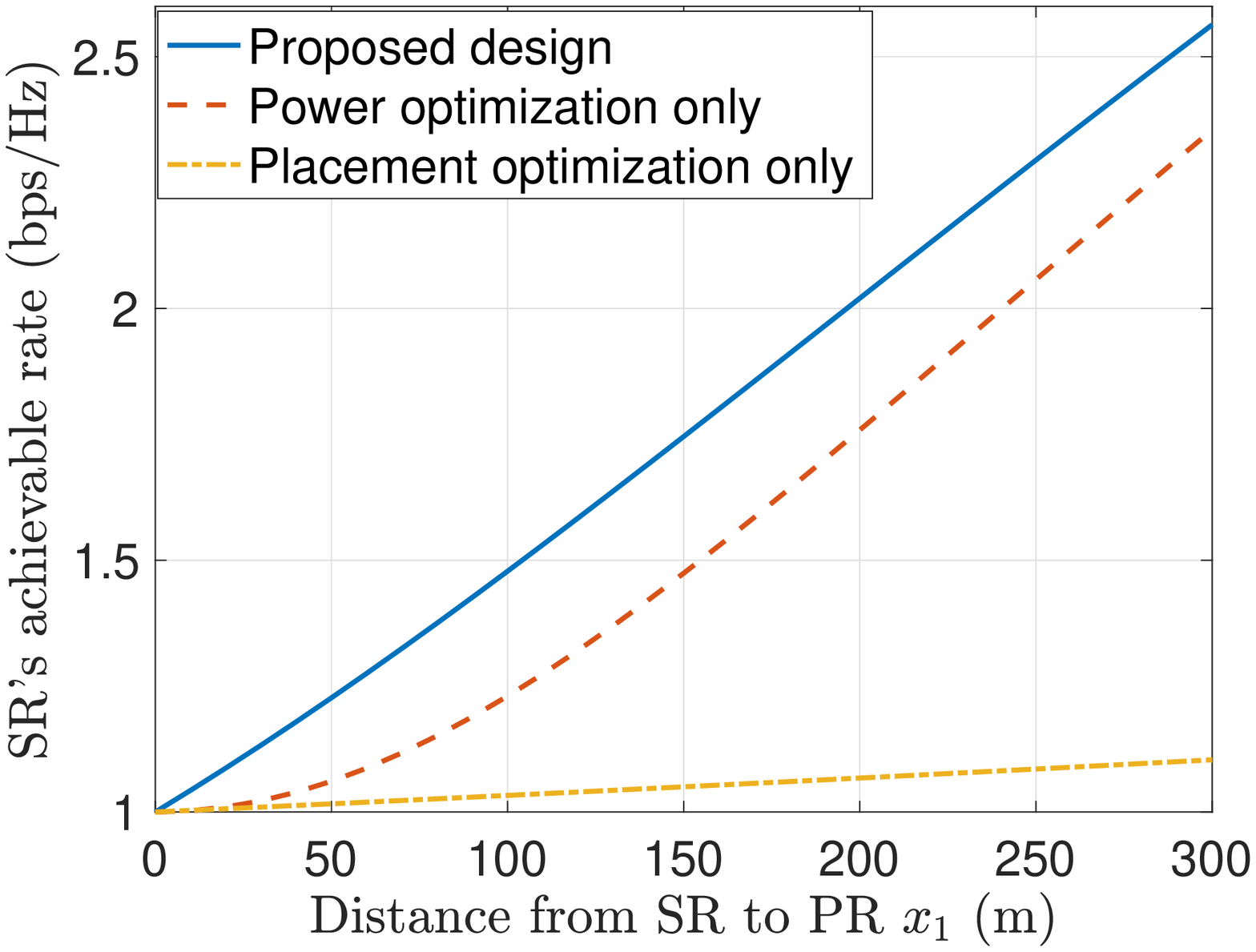}
\caption{SR's achievable rate versus the distance from SR to PR with $K=1$ PR.}
\label{performance_gain}
\end{minipage}
\hfill
\begin{minipage}[t]{0.48\linewidth}
\centering
\includegraphics[width=7cm]{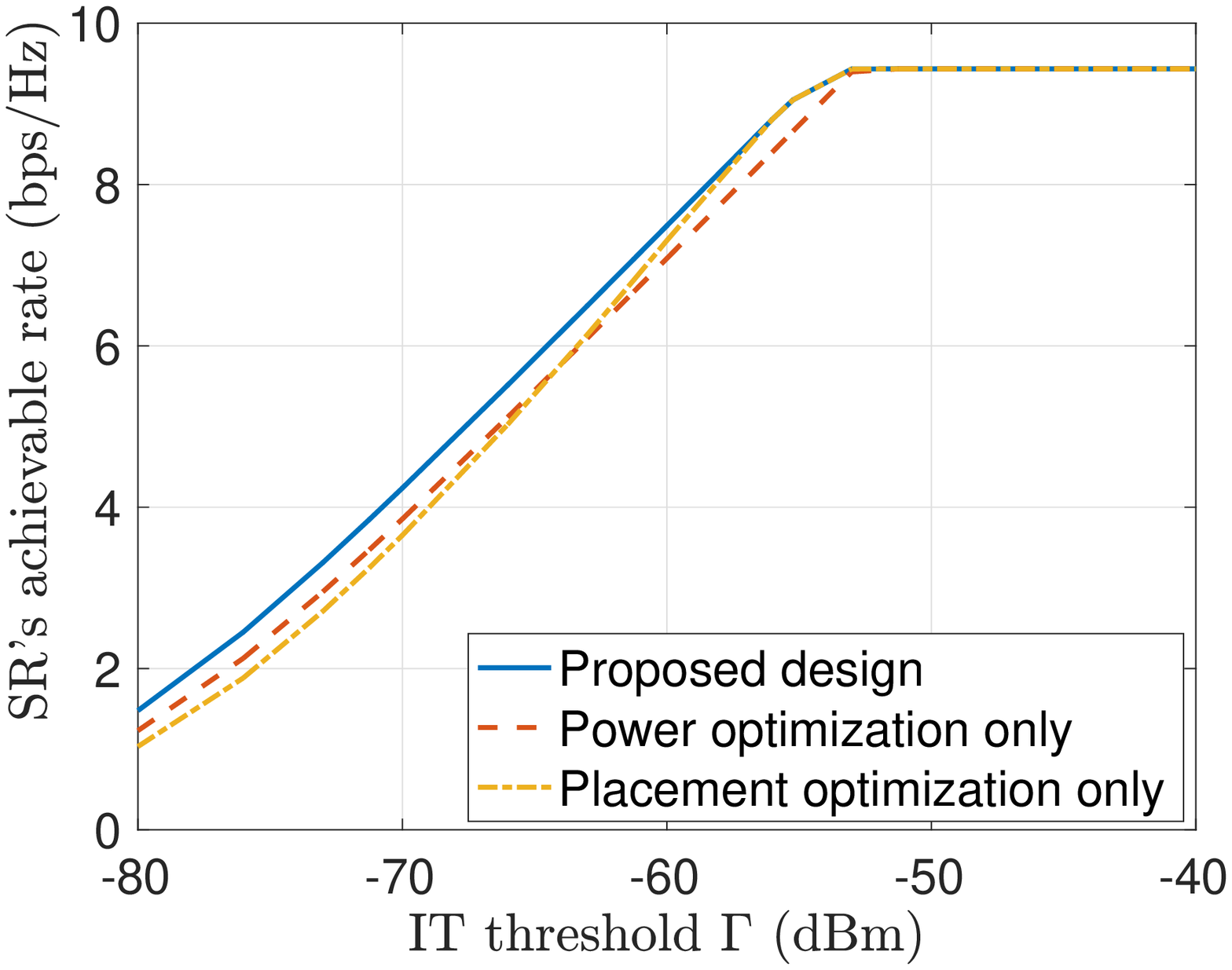}
\caption{SR's achievable rate versus the IT threshold with $K=1$ PR.}
\label{rate_it}
\end{minipage}

\begin{minipage}[t]{0.48\linewidth}
\centering
\includegraphics[width=7cm]{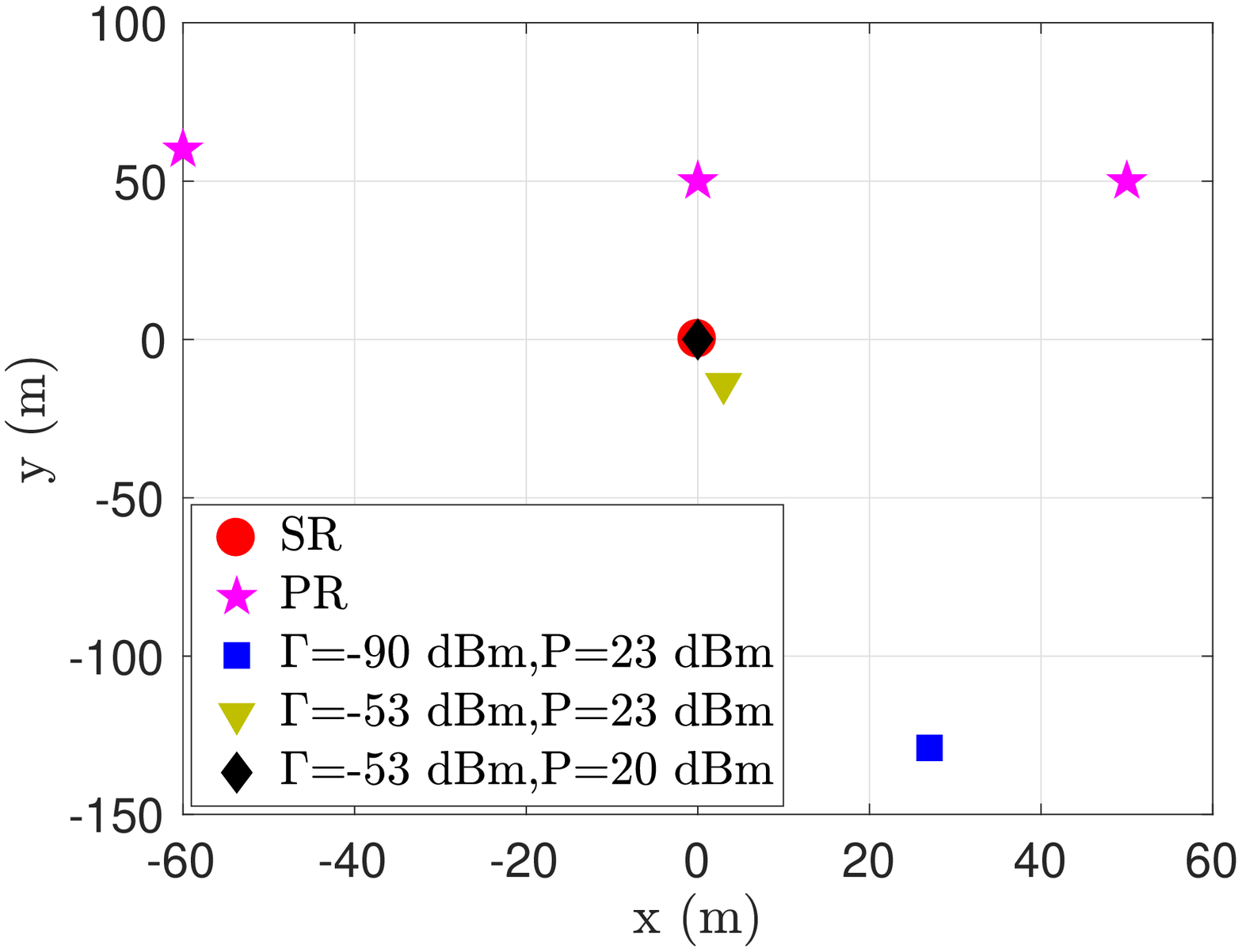}
\caption{UAV's optimal horizontal locations with $K=3$ PRs.}
\label{hover}
\end{minipage}
\hfill
\begin{minipage}[t]{0.48\linewidth}
\centering
\includegraphics[width=7cm]{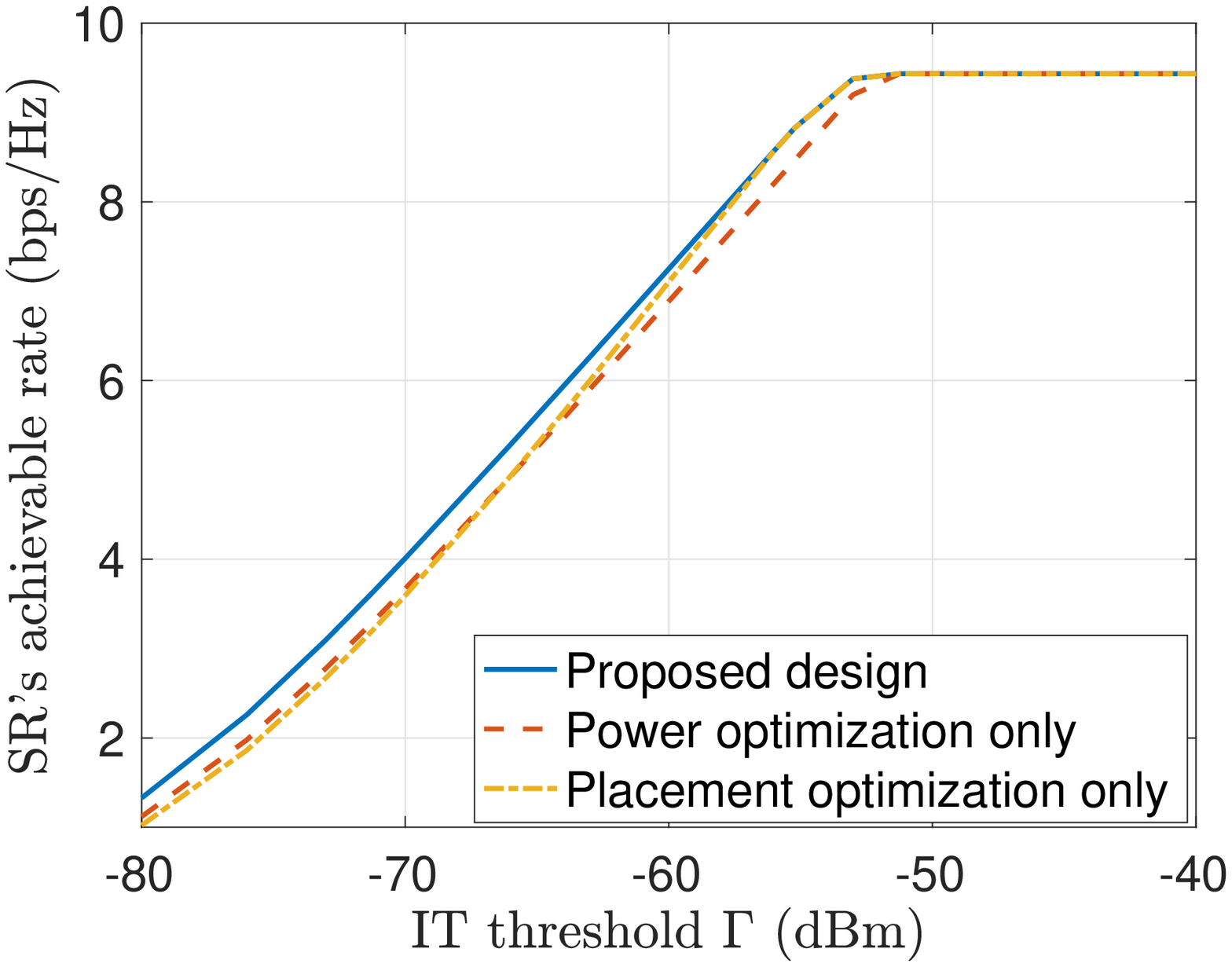}
\caption{SR's achievable rate versus the IT threshold with $K=3$ PRs.}
\label{rate_it_general}
\end{minipage}
\end{figure*}

\vspace{-10pt}\subsection{Quasi-Stationary UAV Scenario}
In this subsection, we evaluate the performance of our proposed optimal solution to (P1) for the quasi-stationary UAV scenario, as compared to the following two benchmark schemes.
\begin{itemize}
\item{\bf Power optimization only:} The UAV is placed exactly above the SR with the lowest altitude, i.e., $(\mv q,z)=(0,0,H_{\min})$. In this case, analogous to (\ref{poweroptimization}), the UAV's optimal transmit power is obtained as $\bar{p}^{*}=\min(P,\min\limits_{k\in\mathcal K}\frac{\Gamma}{\beta_{0}}(\mv w_{k}^{2}+H_{\min}^{2})^{\alpha/2}).$
\item{\bf Placement optimization only:} The UAV optimizes its location $(\mv q,z)$ with the maximum transmit power $P$ used, i.e., $p=P$. This corresponds to solving problem (P1) under given $p=P$, for which the optimal solution can be obtained by applying the SDR technique, which is similar as in Section $\text{\uppercase\expandafter{\romannumeral3}}$-B.
\end{itemize}

First, we consider the case with $K=1$ PR with the PR located at $(x_{1},0,0)$ with $x_{1}\geq 0$. Fig. \ref{performance_gain} shows the SR's achievable rate versus the distance $x_{1}$ from the SR to the PR with $\Gamma=\text{--80 dBm}$ and $P=\text{23 dBm}$. It is observed that as the distance $x_{1}$ from the SR to the PR increases, the achievable rates by all schemes increase. This is due to the fact that when the PR is away from the SR, the IT constraint at the PR becomes less stringent. It is also observed that when the PR is located very close to the SR (i.e., $x_{1}\to 0$), the performance gap between the proposed design and the two benchmark schemes is negligible. This is because in this case, the SR's received signal power is fundamentally limited by the PR's IT constraint, thus leading to the comparable performance for the three schemes. By contrast, as the distance $x_{1}$ increases, the performance gap between the proposed and benchmark schemes is observed to be enlarged.

Fig. \ref{rate_it} shows the SR's achievable rate versus the IT threshold $\Gamma$ with $\mv w_{1}=\text{(100 m, 0 m)}$ and $P=\text{23 dBm}$. It is observed that when $\Gamma$ is sufficiently large (e.g., $\Gamma\geq \text{--53 dBm}$), all schemes achieve the same rate performance. This is because in this case, the transmit power constraint dominates the IT constraints, and thus the three schemes become equivalent. However, when $\Gamma$ becomes smaller (e.g., $\Gamma< \text{--53 dBm}$), our proposed design is observed to outperform the two benchmark schemes. In particular, when $\Gamma$ is sufficiently small (e.g., $\Gamma=\text{--80 dBm}$), the SR's achievable rate by our proposed design is approximately $20\%$ more than that by the scheme with power optimization only, and $40\%$ more than that by the scheme with placement optimization only (as also shown in Fig. \ref{performance_gain} more clearly).
\begin{figure}
\centering
\includegraphics[width=7cm]{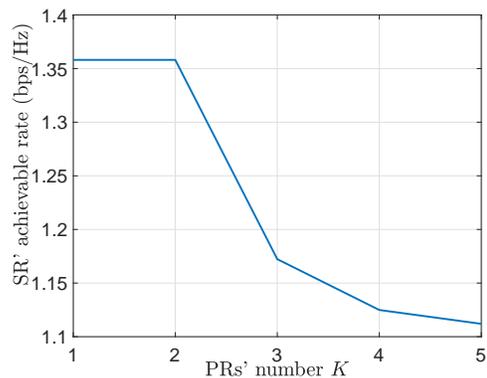}
\caption{SR's achievable rate of the proposed design versus the number of PRs $K$.}
\label{ratek}
\vspace{-10pt}\end{figure}

\begin{figure*}
\begin{minipage}[t]{0.48\linewidth}
\centering
\includegraphics[width=7cm]{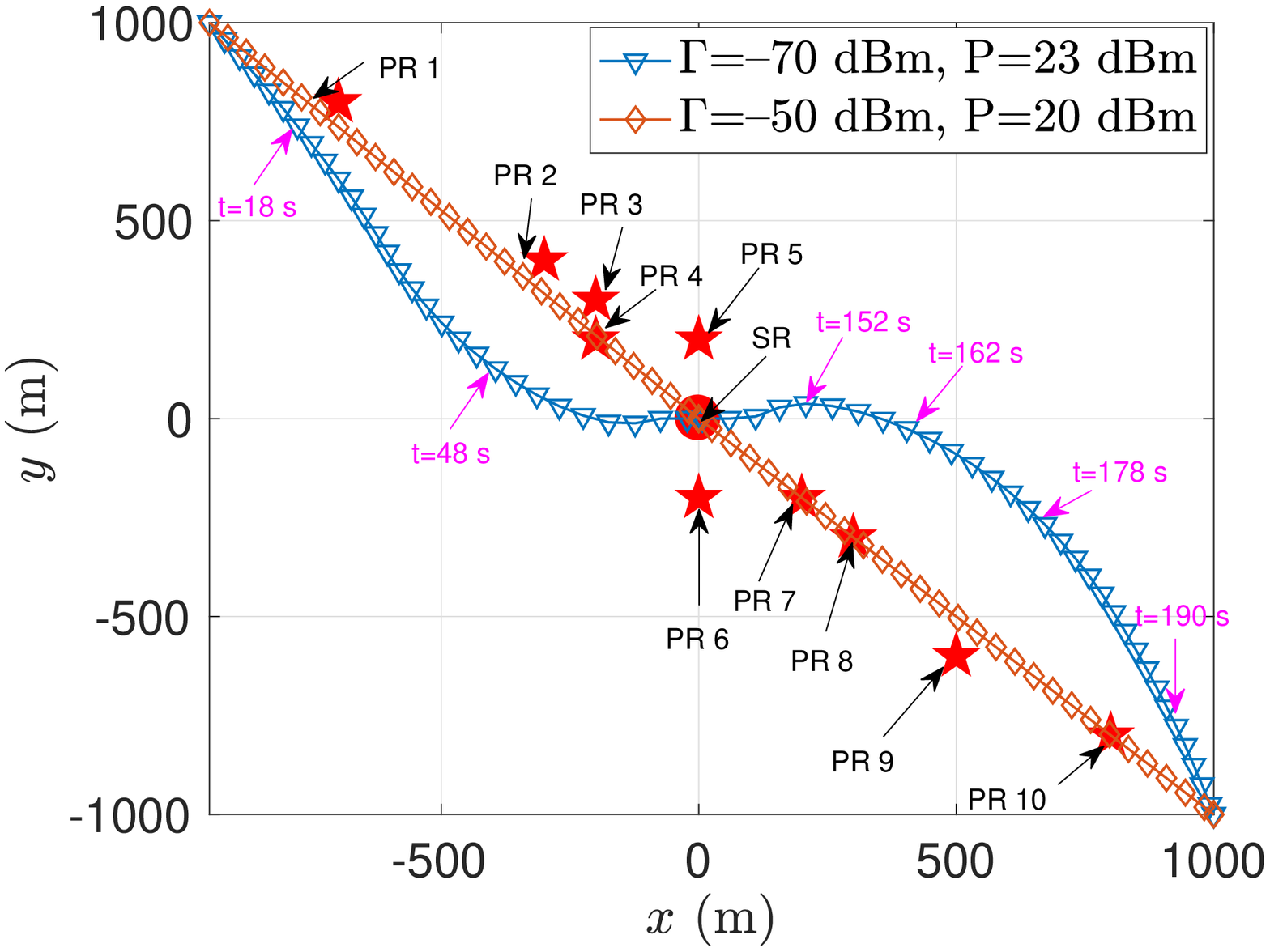}
\caption{UAV's horizontal trajectories.}
\label{trajectoryxy}
\end{minipage}
\hfill
\begin{minipage}[t]{0.48\linewidth}
\centering
\includegraphics[width=7cm]{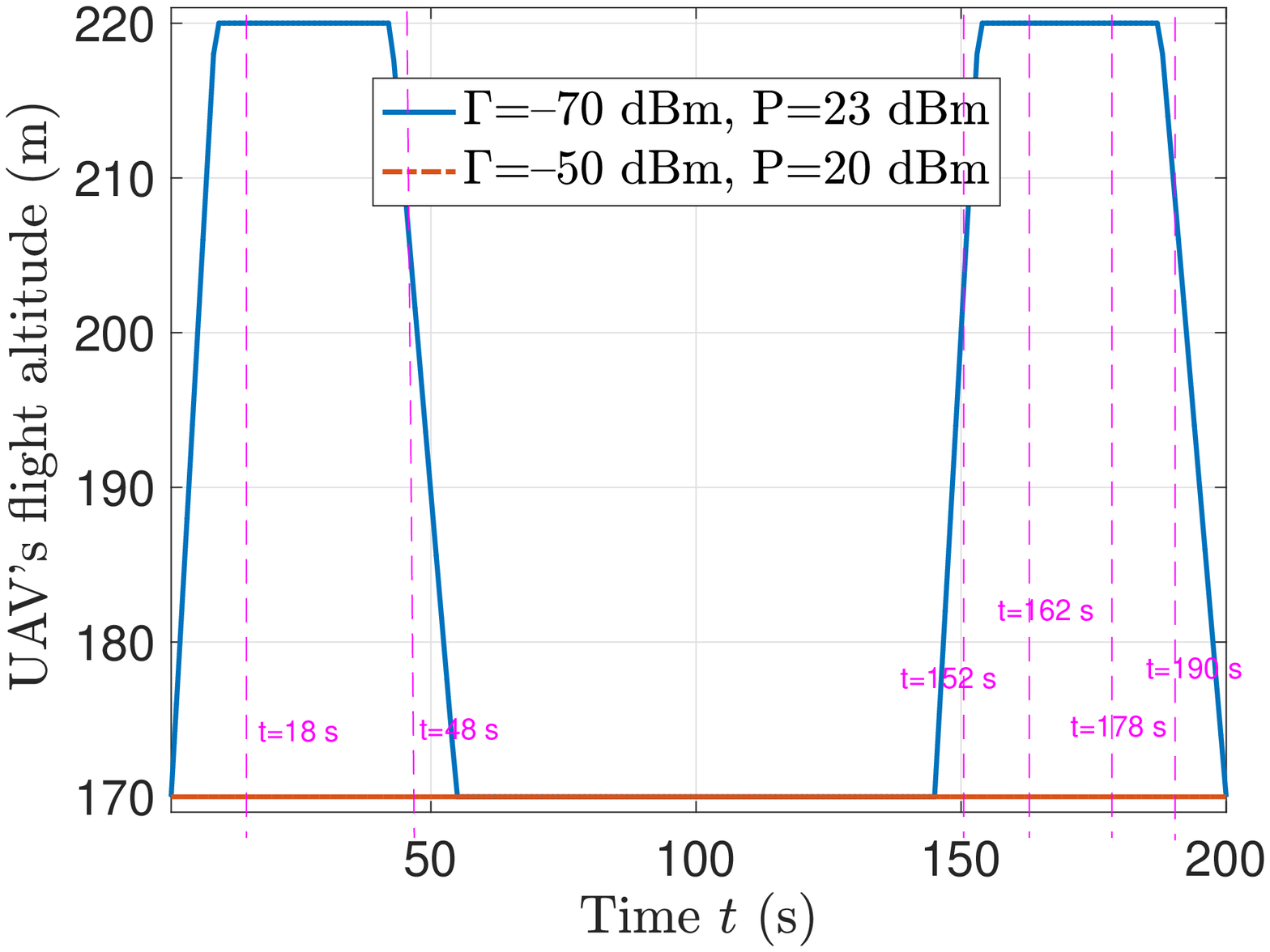}
\caption{UAV's flight altitudes over time.}
\label{trajectoryz}
\end{minipage}

\begin{minipage}[t]{0.48\linewidth}
\centering
\includegraphics[width=7.2cm]{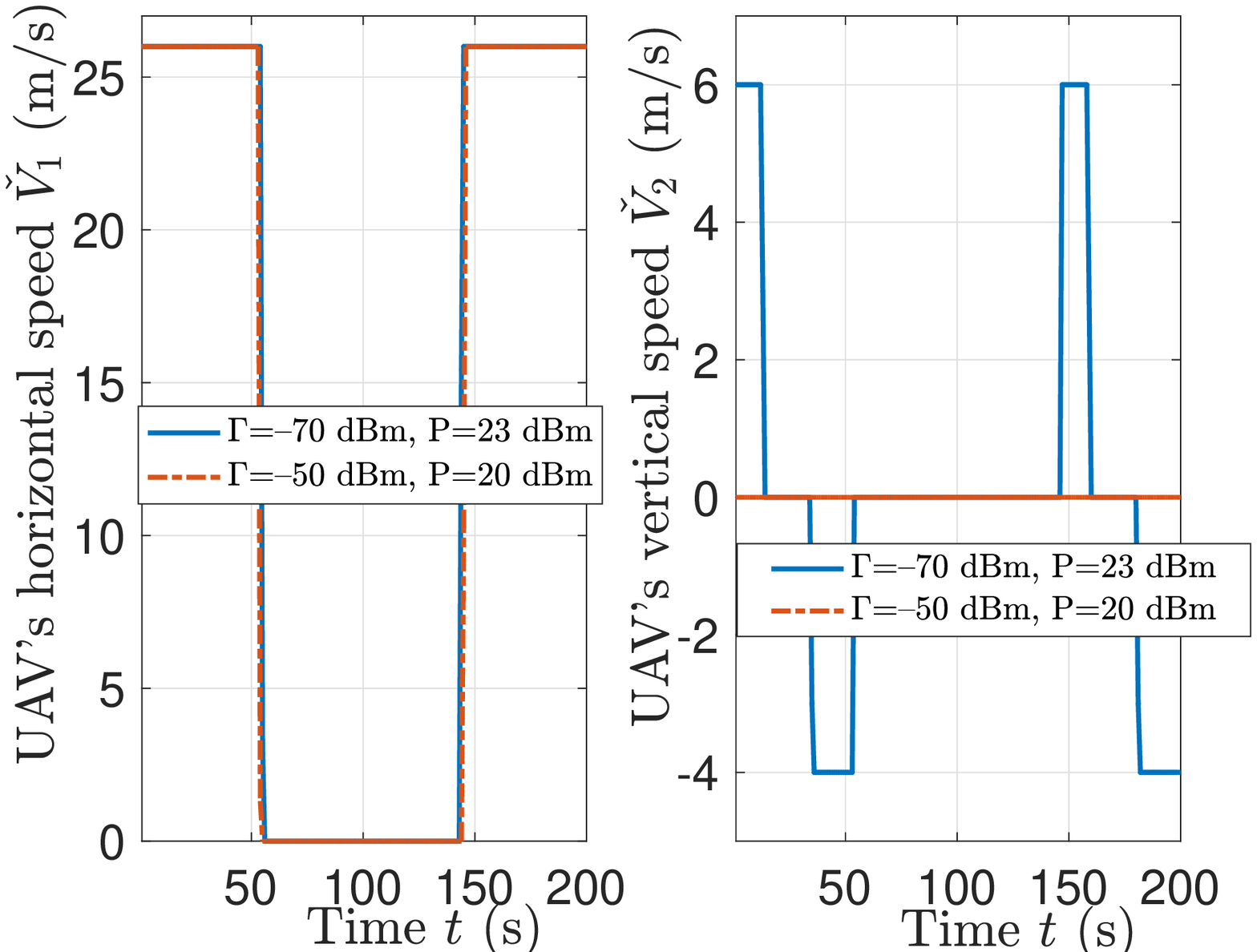}
\caption{UAV's horizontal and vertical speeds over time.}
\label{speed}
\end{minipage}
\hfill
\begin{minipage}[t]{0.48\linewidth}
\centering
\includegraphics[width=7.2cm]{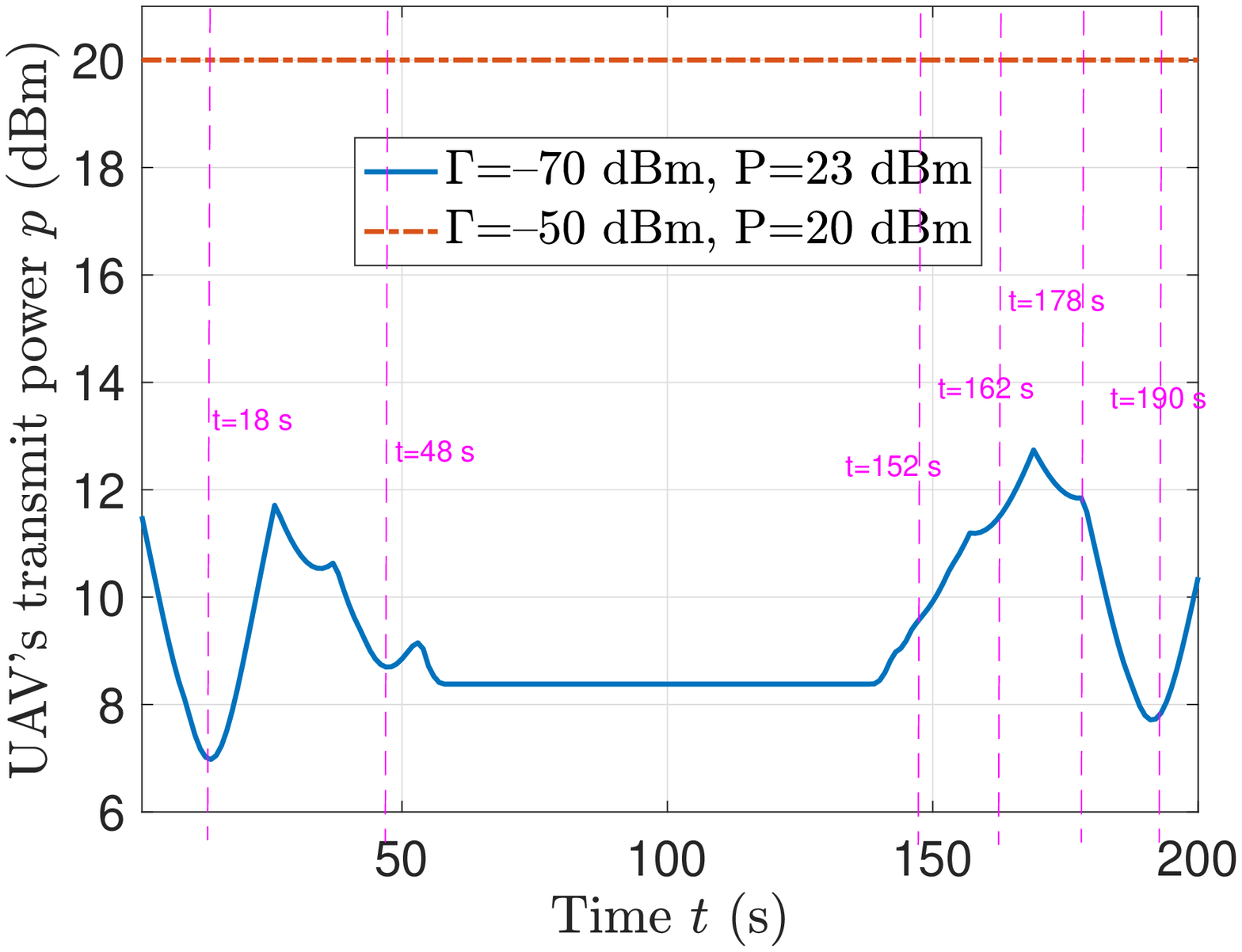}
\caption{UAV's transmit powers over time.}
\label{power}
\end{minipage}
\vspace{-10pt}
\end{figure*}
Next, we consider the setup with $K=3$ PRs as shown in Fig. \ref{hover}\footnote{Note that our proposed SDR-based solution is applicable to any locations of PRs. In Fig. \ref{hover}, we consider that the PRs are all located at the same side of the SR, only for the purpose of showing the impacts of $\Gamma$ and $P$ on the UAV's optimal location.}. Based on Proposition \ref{hmin}, the UAV should always be placed at the lowest altitude with $z^{\star}=H_{\min}=\text{170 m}$. Therefore, for simplicity, only the optimized horizontal locations are shown in Fig. \ref{hover}. It is observed that when the UAV's maximum transmit power $P$ increases and/or the IT threshold $\Gamma$ decreases, the UAV needs to move further away from the PRs, so as to meet the interference constraints at the PRs. Fig. \ref{rate_it_general} shows the SR's achievable rate versus the IT threshold $\Gamma$ with $P=\text{23 dBm}$. Similar observations can be made as in Fig. \ref{rate_it}, where the performance gains over the two benchmark schemes are more significant with smaller values of $\Gamma$.

Furthermore, Fig. \ref{ratek} shows the SR's achievable rate of the proposed design versus the number of PRs, $K$. For each $K$, we randomly generate the PRs' locations in an area of $200\times 200~\text{m}^{2}$. The results are obtained by averaging over 100 random realizations, where we set $\Gamma=\text{--90 dBm}$ and $P=23~\text{dBm}$. It is observed that as $K$ increases, the SR's achievable rate is non-increasing. This is because as $K$ increases, the number of IT constraints increases, thus making the feasibility region of (P1) smaller. As a result, the SR's achievable rate may decrease.

\vspace{-10pt}\subsection{Mobile UAV Scenario}
In this subsection, we evaluate the performance of our proposed solution to (P2) under the mobile UAV scenario. Unless otherwise stated, we assume that there are $K=10$ ground PRs distributed in a 2D area of $2\times 2~\text{km}^{2}$, as shown in Fig. \ref{trajectoryxy}. The speed limits for the UAV are set according to DJI's Inspire 2 drones \cite{dajiang}, i.e., $\hat{V}_{H}=\text{26 m/s}$, $\hat{V}_{A}=\text{6 m/s}$ and $\hat{V}_{D}=\text{4 m/s}$. The UAV's initial and final locations are set as $\text{(--950 m, 1000 m, 170 m)}$ and $\text{(1000 m, --1000 m, 170 m)}$, respectively. In this case, the UAV's minimum flight duration is $T_{\min}=\text{107 s}$.

Figs. \ref{trajectoryxy} and \ref{trajectoryz} show the UAV's horizontal locations and flight altitudes over time by the proposed design, under different values of $P$ and $\Gamma$ with the communication duration set as $T=\text{200 s}$. Figs. \ref{speed} and \ref{power} show the corresponding UAV's flying speeds and the optimized transmit powers over time, respectively. It is observed that when $\Gamma=\text{--50 dBm}$ and $P=\text{20 dBm}$, the UAV flies simply following its initial FHF trajectory, along which the UAV always stays at the minimum altitude and transmits with the full power $P$. This is consistent with Proposition \ref{pzlemma}, which shows that when the maximum transmit power $P$ is sufficiently small and/or the IT threshold $\Gamma$ is sufficiently large, the UAV should stay at the minimum altitude and transmit with the maximum power. However, when $\Gamma$ decreases and $P$ increases in the case of $\Gamma=\text{--70 dBm}$ and $P=\text{23 dBm}$, the UAV trajectory is observed to deviate from the initial FHF trajectory. In particular, when the UAV approaches PRs 1--4 and PRs 7--10 (at time instants $t=\text{18 s}$, $t=\text{48 s}$, $t=\text{152 s}$, $t=\text{162 s}$, $t=\text{178 s}$ and $t=\text{190 s}$, respectively), it increases the flight altitude and reduces transmit power, in order to meet the IT constraint at the nearest PR. Notice that this observation is also consistent with Proposition \ref{pzlemma} and Remark \ref{compare}, which reveals that the UAV should increase its altitude to maximize the cognitive communication rate when it moves closer to some PRs than the SR. Furthermore, it is observed that the UAV hovers above a point closer to the SR than all PRs at the lowest altitude for a certain period of time to take advantage of the favorable communication channel with the SR for enhancing the SR's achievable rate. This is consistent with Proposition \ref{hmin}.

\begin{figure}
\centering
\includegraphics[width=7cm]{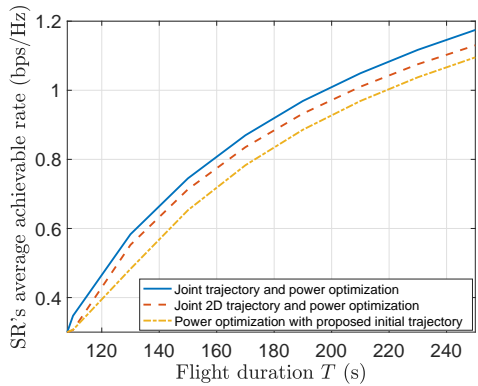}
\caption{SR's average achievable rate versus the flight duration $T$.}
\label{rate}
\vspace{-10pt}\end{figure}

Finally, Fig. \ref{rate} shows the SR's average achievable rate by the proposed design with $\Gamma=\text{--80 dBm}$ and $P=\text{23 dBm}$, as compared to the following two benchmark schemes.
\begin{itemize}
\item[$\bullet$] {\bf Joint 2D trajectory and power optimization:} The UAV jointly optimizes its 2D trajectory $\{\mv q[n]\}$ and transmit power $\{p[n]\}$, where the flight altitude is fixed as its minimum flight altitude. This corresponds to solving problem (P2) under given $z[n]=H_{\min}$, $\forall n\in\mathcal N$.
\item[$\bullet$] {\bf Power optimization with proposed initial trajectory:} The UAV sets its trajectory as the proposed initial trajectory, as given in Remark \ref{initial}. Under this trajectory, the UAV optimizes its power allocation based on (\ref{poweroptimization}).
\end{itemize}
From Fig. \ref{rate}, it is observed that as the flight duration $T$ increases, the average achievable rates by all the three schemes increase. This is because for all cases with adaptive trajectory design, the UAV in general stays longer close to the SR when $T$ increases, leading to a better channel condition on average and thus a higher average achievable rate. It is also observed that the proposed joint 3D trajectory and power design outperforms its 2D counterpart. This is because in the proposed design, the UAV can adjust its altitude more freely to control the co-channel interference, especially when the UAV is close to the PRs. This is consistent with our observations in Remark \ref{compare} and validates the importance of 3D trajectory design with altitude control.

\section{Concluding Remarks}
This paper studied a new spectrum sharing scenario for UAV-to-ground communications, where a cognitive/secondary UAV transmitter communicates with a ground SR, in the presence of co-channel primary terrestrial wireless communication links. We exploited the UAV's 3D mobility to improve the cognitive communication rate performance under two scenarios of {\it quasi-stationary} and {\it mobile} UAVs, respectively. For both scenarios, we proposed efficient algorithms to obtain high-quality solutions to the joint UAV maneuver and power control optimization problems. It was shown via simulations that the proposed designs with joint 3D placement/trajectory and power control optimization significantly outperform benchmark schemes without such a joint design or with only 2D optimization. Due to the space limitation, there are other important issues that remain unaddressed yet in this paper, which are discussed in the following to motivate future work.
\begin{itemize}
\item This paper considered the offline UAV maneuver design by assuming that the UAV perfectly knows the channel parameters in advance. Such offline design, however, may lead to sub-optimal performance in real-time implementation. This is because the deterministic LoS channel (or even stochastic channels) model may mismatch with realistic radio propagation environments, due to the unevenly distributed obstacles (such as buildings and trees) around. How to optimize UAV maneuver based on the actual channel is thus an important problem to be tackled in future work. In this case, a promising solution is by using the radio map technique \cite{radiomap} to obtain the location-dependent channel knowledge offline, or adopting reinforcement learning to adapt the UAV maneuver to the actual channel in real time. Accordingly, our proposed solutions based on the {\it a-priori} known LoS channel model can not only provide a performance upper bound for practical maneuver design, but also serve as an initial input for the online design.
\item This paper considered the basic setup with one UAV and one SR. In practice, there may exist multiple coexisting SRs and UAVs within the same network. In the case with one UAV communicating with multiple SRs, the UAV needs to properly schedule its transmission to the multiple SRs based on the adopted multiple access scheme (e.g., time division multiple access (TDMA), orthogonal frequency division multiple access (OFDMA), or non-orthogonal multiple-access (NOMA)). How to jointly design the UAV maneuver, SRs' scheduling and resource allocation is an interesting problem worth pursuing in future work. Furthermore, when there are multiple UAVs, the co-channel interference from UAVs to SRs becomes a new issue to be dealt with. For instance, these UAVs can jointly design their maneuvers and power allocations to maximize their weighted sum rates, while ensuring that their caused aggregate interference power at each PR does not exceed the prescribed IT constraint. Moreover, different SRs may use the coordinated multi-point (CoMP) technique to jointly decode the messages from multiple UAVs for better mitigating or even utilizing the strong co-channel A2G interference. These problems are worthy of more in-depth investigation in future work.
\end{itemize}

\appendix
\subsection{Proof of Proposition \ref{pzlemma}}\label{pzlemmaproof}
First, we recast constraint (\ref{pzIT1}) as $\hat{p}\leq \frac{\hat{\Gamma}}{\hat{\beta}_{0}}(z^{2}+\Vert\mv q-\mv w_{\tilde{k}(\bm q)}\Vert^{2})$. Obviously, at least one of the constraints (\ref{staticpower}) and (\ref{pzIT1}) must be tight at the optimality of (P3). Notice that $\frac{\hat{\Gamma}}{\hat{\beta}_{0}}(H_{\min}^{2}+\Vert\mv q-\mv w_{\tilde{k}(\bm q)}\Vert^{2})\leq \frac{\hat{\Gamma}}{\hat{\beta}_{0}}(z^{2}+\Vert\mv q-\mv w_{\tilde{k}(\bm q)}\Vert^{2})\leq \frac{\hat{\Gamma}}{\hat{\beta}_{0}}(H_{\max}^{2}+\Vert\mv q-\mv w_{\tilde{k}(\bm q)}\Vert^{2})$ due to $H_{\min}\leq z\leq H_{\max}$. Based on this, we consider the following three cases to obtain the optimal solution to (P3).

If $\hat{p}>\frac{\hat{\Gamma}}{\hat{\beta}_{0}}(H_{\max}^{2}+\Vert\mv q-\mv w_{\tilde{k}(\bm q)}\Vert^{2})$, then only constraint (\ref{pzIT1}) is tight at the optimality of (P3), and thus we have
\begin{align}\label{pq}
\hat{p}(\mv q)=\frac{\hat{\Gamma}}{\hat{\beta}_{0}}(z^{2}+\Vert\mv q-\mv w_{\tilde{k}(\bm q)}\Vert^{2}).
\end{align}
The corresponding objective value of (P3) is expressed as
\begin{align}\label{fz}
f(z)=\frac{\hat{\Gamma}}{\hat{\beta}_{0}}\frac{\Vert\mv q-\mv w_{\tilde{k}(\bm q)}\Vert^{2}+z^{2}}{\Vert\mv q\Vert+z^{2}}.
\end{align}
Then, we consider the following three cases to obtain the maximum value of $f(z)$.
\begin{itemize}
\item In Case 1 (i.e., $\Vert\mv q-\mv w_{\tilde{k}(\bm q)}\Vert<\Vert\mv q\Vert$), it can be verified that $f(z)$ monotonically increases with $z\in[H_{\min},H_{\max}]$, and thus we have $z^*(\mv q)=H_{\max}$. By substituting this into (\ref{pq}), we have $\hat{p}^*(\mv q)=\frac{\hat{\Gamma}}{\hat{\beta}_{0}}(\Vert\mv q-\mv w_{\tilde{k}(\bm q)}\Vert^{2}+H_{\max}^{2})$.
\item In Case 2 (i.e., $\Vert\mv q\Vert<\Vert\mv q-\mv w_{\tilde{k}(\bm q)}\Vert$), it can be verified that  $f(z)$ monotonically decreases with $z\in[H_{\min},H_{\max}]$, and thus we have $z^{*}(\mv q)=H_{\min}$. By substituting this into (\ref{pq}), we have $\hat{p}^{*}(\mv q)=\frac{\hat{\Gamma}}{\hat{\beta}_{0}}(\Vert\mv q-\mv w_{\tilde{k}(\bm q)}\Vert^{2}+H_{\min}^{2})$.
\item In Case 3 (i.e., $\Vert\mv q\Vert=\Vert\mv q-\mv w_{\tilde{k}(\bm q)}\Vert$), we have $f(z)=\hat{\Gamma}/\hat{\beta}_{0}$, which is regardless of the UAV's flight altitude $z$. Thus, the optimal flight altitude $z^{*}(\mv q)$ can be an arbitrary value within the interval $[H_{\min},H_{\max}]$. By substituting this into (\ref{pq}), we have $\hat{p}^{*}(\mv q)=\frac{\hat{\Gamma}}{\hat{\beta}_{0}}(z^{*2}(\mv q)+\Vert\mv q-\mv w_{\tilde{k}(\bm q)}\Vert^{2})$.
\end{itemize}

If $\hat{p}<\frac{\hat{\Gamma}}{\hat{\beta}_{0}}(H_{\min}^{2}+\Vert\mv q-\mv w_{\tilde{k}(\bm q)}\Vert^{2})$, then only the power constraint (\ref{staticpower}) is tight at the optimality of (P3). Thus, the UAV can be placed at the lowest altitude and transmit at the maximum power to maximize the received power at the SR, i.e., $\hat{p}^{*}(\mv q)=P$ and $z^{*}(\mv q)=H_{\min}$.

Finally, if $\frac{\hat{\Gamma}}{\hat{\beta}_{0}}(\Vert\mv q-\mv w_{\tilde{k}(\bm q)}\Vert^{2}+H_{\min}^{2})<\hat{p}<\frac{\hat{\Gamma}}{\hat{\beta}_{0}}(\Vert\mv q-\mv w_{\tilde{k}(\bm q)}\Vert^{2}+H_{\max}^{2})$, it follows that constraint (\ref{pzIT1}) must be tight at the optimality of (P3), since otherwise we can decrease the UAV's altitude and/or increase the UAV's transmit power to increase the SR's achievable rate, without violating the PR's IT constraint. Therefore, we still have $\hat{p}(\mv q)$ in (\ref{pq}) and $f(z)$ in (\ref{fz}). By substituting (\ref{pq}) into the power constraint (\ref{staticpower}), we have $z\leq \sqrt{\frac{\hat{\beta}_{0}\hat{p}}{\hat{\Gamma}}-\Vert\mv q-\mv w_{\tilde{k}(\bm q)}\Vert^{2}}$. With $\frac{\hat{\Gamma}}{\hat{\beta}_{0}}(\Vert\mv q-\mv w_{\tilde{k}(\bm q)}\Vert^{2}+H_{\min}^{2})<\hat{p}<\frac{\hat{\Gamma}}{\hat{\beta}_{0}}(\Vert\mv q-\mv w_{\tilde{k}(\bm q)}\Vert^{2}+H_{\max}^{2})$, it follows that $H_{\min}<\sqrt{\frac{\hat{\beta}_{0}\hat{p}}{\hat{\Gamma}}-\Vert\mv q-\mv w_{\tilde{k}(\bm q)}\Vert^{2}}<H_{\max}$. Thus, we can obtain $\small{H_{\min}\leq z\leq \sqrt{\frac{\hat{\beta}_{0}\hat{p}}{\hat{\Gamma}}-\Vert\mv q-\mv w_{\tilde{k}(\bm q)}\Vert^{2}}}$. Next, by checking the monotonicity of $f(z)$ (similarly as the case with $\small{\hat{P}}>\frac{\hat{\Gamma}}{\hat{\beta}_{0}}(H_{\max}^{2}+\Vert\bm q-\bm w_{\tilde{k}(\bm q)}\Vert^{2})$), we can obtain the following results:
\begin{itemize}
\item In Case 1 (i.e., $\Vert\mv q-\mv w_{\tilde{k}(\bm q)}\Vert<\Vert\mv q\Vert$), $z^*(\mv q)=\sqrt{\frac{\hat{\beta}_{0}\hat{p}}{\hat{\Gamma}}-\Vert\mv q-\mv w_{\tilde{k}(\bm q)}\Vert^{2}}$ and $\hat{p}^*(\mv q)=\hat{p}$.
\item In Case 2 (i.e., $\Vert\mv q\Vert<\Vert\mv q-\mv w_{\tilde{k}(\bm q)}\Vert$), $z^*(\mv q)=H_{\min}$ and $\hat{p}^{*}(\mv q)=\frac{\hat{\Gamma}}{\hat{\beta}_{0}}(\Vert\mv q-\mv w_{\tilde{k}(\bm q)}\Vert^{2}+H_{\min}^{2})$.
\item In Case 3 (i.e., $\Vert\mv q\Vert=\Vert\mv q-\mv w_{\tilde{k}(\bm q)}\Vert$), $z^{*}(\mv q)$ is an arbitrary value within the interval $[H_{\min},\sqrt{\frac{\hat{\beta}_{0}\hat{p}}{\hat{\Gamma}}-\Vert\mv q-\mv w_{\tilde{k}(\bm q)}\Vert^{2}}]$, and $\hat{p}^{*}(\mv q)=\frac{\hat{\Gamma}}{\hat{\beta}_{0}}(z^{*2}(\mv q)+\Vert\mv q-\mv w_{\tilde{k}(\bm q)}\Vert^{2})$
\end{itemize}
By combining all the results above and with some manipulation, this proposition is proved.

\vspace{-10pt}\subsection{Proof of Lemma \ref{probabilityp}}\label{probability}
First, we define $p_{1}\triangleq\frac{\hat{\Gamma}}{\hat{\beta}_{0}}(\min\limits_{k\in\mathcal K}\Vert\mv w_{k}\Vert^{2}+H_{\min}^{2})$, which denotes the maximally allowable value of $\hat{p}$ for IT constraints (\ref{staticIT}) to be feasible, in the case when the UAV is located exactly above the SR at $(0,0, H_{\min})$. Notice that if $\hat{p}$ is sufficiently low with $\hat{p}\leq p_{1}$, the UAV can be placed above the SR at the lowest altitude. Then Lemma \ref{probabilityp} holds accordingly. As a result, it only remains to consider the case with $\hat{p}>p_{1}$, for which we can prove Lemma \ref{probabilityp} by contradiction. Specifically, suppose that, at the optimal solution, the UAV is placed closer to a PR $k\in\mathcal K$ than the SR with $\Vert\mv q^{\star}-\mv w_{k}\Vert< \Vert\mv q^{\star}\Vert,~\forall k\in\mathcal K$. By combining constraints (\ref{staticpower}) and (\ref{pzIT1}), we have $\hat{p}^{\star}\leq \min(\hat{p},\frac{\hat{\Gamma}}{\hat{\beta}_{0}}(\Vert\mv q^{\star}-\mv w_{k}\Vert^{2}+z^{2})),~\forall k\in\mathcal K$, and the corresponding objective value of (P1.1) can be obtained as
\begin{align}\xi_{1}&\leq \frac{\min(\hat{p},\frac{\hat{\Gamma}}{\hat{\beta}_{0}}(\Vert\bm q^{\star}-\bm w_{k}\Vert^{2}+z^{2}))}{\Vert\bm q^{\star}\Vert^{2}+z^{2}}\nonumber\\
&\leq \frac{\hat{\Gamma}}{\hat{\beta}_{0}}\frac{\Vert\bm q^{\star}-\bm w_{k}\Vert^{2}+z^{2}}{\Vert\bm q^{\star}\Vert^{2}+z^{2}}\triangleq \xi_{2},~\forall k\in\mathcal K.\nonumber
\end{align}
 Due to $\Vert\mv q^{\star}-\mv w_{k}\Vert<\Vert\mv q^{\star}\Vert,~\forall k\in\mathcal K,$ it follows that $\xi_{1}\leq \xi_{2}<\hat{\Gamma}/\hat{\beta}_{0}$. Next, it is easy to verify that $(\mv q,z,p)=(\mv 0,H_{\min},p_{1})$ is a feasible solution to (P1) (i.e., the UAV hovers right above the SR at the lowest altitude). The corresponding objective value of (P1.1) can be obtained as $\xi_{3}=\frac{\hat{\Gamma}}{\hat{\beta}_{0}}\frac{\min_{k\in\mathcal K}\Vert\bm w_{k}\Vert^{2}+H_{\min}^{2}}{H_{\min}^{2}}$. It is evident that  $\xi_{3}>\hat{\Gamma}/\hat{\beta}_{0}> \xi_{2}\geq \xi_{1}$, which contradicts the optimality of $\mv q^{\star}$.
Therefore, this lemma is proved.

\vspace{-10pt}\subsection{Proof of Proposition \ref{rankone}}\label{rankoneproof}
Suppose that the optimal solution to problem (P4.3) is denoted by $\hat{p}^{\star}$, $\tau^{\star}$, and $\mv S^{\star}$. Then we construct the following problem:
\begin{align}
\text{(P4.5):}~\max\limits_{E,\bm S}~&E\nonumber\\
\text{s.t.}~&\text{Tr}(\mv B_{k}\mv S)\geq E, ~\forall k\in\mathcal K,\label{p26it}\\
&\text{Tr}(\mv A\mv S)\leq \frac{\hat{p}^{\star}}{\tau^{\star}}-H_{\min}^{2},\label{p25t}\\
&\text{Tr}(\mv C\mv S)=1,\label{p23m1}\\
&\mv S\succeq \mv 0.\label{p23s01}
\end{align}
It is evident that problem (P4.5) and problem (P4.3) have the same optimal solution of $\mv S$. Therefore, $\mv S^{\star}$ is also optimal for (P4.5). Hence, to prove this proposition, we only need to show that when the $K$ PRs are located at the same side of the SR, we have $\text{rank}(\mv S^{\star})=1$ for problem (P4.5). Notice that (P4.5) is a convex SDP and satisfies the Slater's condition. Therefore, strong duality holds between problem (P4.5) and its dual problem. Let $\gamma_{k}\geq 0$, $\forall k\in\mathcal K$, $\lambda\geq 0$, and $\mu$ denote the dual variables associated with the constraints in (\ref{p26it}), (\ref{p25t}), and (\ref{p23m1}), respectively. Then the Lagrangian of (P4.5) is expressed as
\begin{align}
&\mathcal L(E,\mv S,\lambda,\{\gamma_{k}\},\mu,\mv G)\nonumber\\
&=\left(1-\sum\limits_{k\in\mathcal K}\gamma_{k}\right)E-\lambda\left(H_{\min}^{2}-\frac{\hat{p}^{\star}}{\tau^{\star}}\right)+\mu+\text{Tr}(\mv G\mv S),\label{L}
\end{align}
where $\mv G=-\lambda\mv A-\mu\mv C+\sum\nolimits_{k\in\mathcal K}\gamma_{k}\mv B_{k}$. Accordingly, the dual problem of (P4.5) is given by
\begin{align}
\text{(D4.5)}~\min\limits_{\lambda\geq 0,\{\gamma_{k}\geq 0\},\mu} &-\lambda\left(H_{\min}^{2}-\frac{\hat{p}^{\star}}{\tau^{\star}}\right)+\mu\nonumber\\
\text{s.t.}~&\mv G\preceq \mv 0,\label{G}\\
&\sum\limits_{k\in\mathcal K}\gamma_{k}=1.\label{gamma1}
\end{align}
Denote the optimal solution of (D4.5) as $\lambda^{\star}$, $\gamma_{k}^{\star}$, $\forall k\in\mathcal K$, and $\mu^{\star}$. Accordingly, the resultant $\mv G^{\star}$ can be explicitly expressed as
$$\small\mv G^{\star}=\begin{bmatrix}\sum\limits_{k\in\mathcal K}\gamma_{k}^{\star}-\lambda^{\star}&0&-\sum\limits_{k\in\mathcal K}\gamma_{k}^{\star}x_{k}\\ 0&\sum\limits_{k\in\mathcal K}\gamma_{k}^{\star}-\lambda^{\star}&-\sum\limits_{k\in\mathcal K}\gamma_{k}^{\star}y_{k}\\-\sum\limits_{k\in\mathcal K}\gamma_{k}^{\star}x_{k}&-\sum\limits_{k\in\mathcal K}\gamma_{k}^{\star}y_{k}&-\mu^{\star}+\sum\limits_{k\in\mathcal K}(\gamma_{k}^{\star}(x_{k}^{2}+y_{k}^{2}))\end{bmatrix}.$$
Then the optimal solution to (P4.5) and (D4.5) should satisfy the complementary slackness condition $\text{Tr}(\mv G^{\star}\mv S^{\star})=0$, or equivalently, $\mv G^{\star}\mv S^{\star}=\mv 0$. Therefore, in order to show $\text{rank}(\mv S^{\star})\leq 1$, we only need to show that $\text{rank}(\mv G^{\star})\geq 2$.

Towards this end, in the following we show that $\sum_{k\in\mathcal K}\gamma_{k}^{\star}-\lambda^{\star}\neq 0$ must hold by contradiction. Suppose that $\sum_{k\in\mathcal K}\gamma_{k}^{\star}-\lambda^{\star}=0$. Then we have
$$\small\mv G^{\star}=\begin{bmatrix}0&0&-\sum\limits_{k\in\mathcal K}\gamma_{k}^{\star}x_{k}\\0&0&-\sum\limits_{k\in\mathcal K}\gamma_{k}^{\star}y_{k}\\-\sum\limits_{k\in\mathcal K}\gamma_{k}^{\star}x_{k}&-\sum\limits_{k\in\mathcal K}\gamma_{k}^{\star}y_{k}&-\mu^{\star}+\sum\limits_{k\in\mathcal K}(\gamma_{k}^{\star}(x_{k}^{2}+y_{k}^{2}))\end{bmatrix}.$$
Denote by $d$ an eigenvalue of the matrix $\mv G^{\star}$. Then, we have $\text{det}(\mv G^{\star}-d\mv I)=0$, which leads to
\begin{align}\label{d}
d\left(-d^{2}+\epsilon_{2} d+\epsilon_{1}\right)=0,
\end{align}
where $\small\epsilon_{1}=\left(\sum_{k\in\mathcal K}\gamma_{k}^{\star}x_{k}\right)^{2}+\left(\sum_{k\in\mathcal K}\gamma_{k}^{\star}y_{k}\right)^{2}$, and $\small\epsilon_{2}=-\mu^{\star}+\sum_{k\in\mathcal K}\left(\gamma_{k}^{\star}\left(x_{k}^{2}+y_{k}^{2}\right)\right)$. Due to the fact that $\gamma_{k}^{\star}\geq 0$, $\forall k\in\mathcal K$, $\sum_{k\in\mathcal K}\gamma_{k}^{\star}=1$, and all PRs are located at the same side of the SR (e.g., $x_{k}\geq 0$, $\forall k\in\mathcal K$, and there exists at least one PR $\bar{k}\in\mathcal K$ with $x_{\bar k}>0$, among the four possible cases), we have $\epsilon_{1}>0$. Therefore, there must exist a positive root to equation (\ref{d}), i.e., the matrix $\mv G^{\star}$ has a positive eigenvalue. This contradicts $\mv G\preceq \mv 0$ in (\ref{G}). Hence, $\sum_{k\in\mathcal K}\gamma_{k}^{\star}-\lambda^{\star}$ must be non-zero.

With $\sum_{k\in\mathcal K}\gamma_{k}^{\star}-\lambda^{\star}\neq 0$, it is easy to show that $\text{rank}(\mv G^{\star})\geq 2$ via some simple elementary transformation. Based on $\mv G^{\star}\mv S^{\star}=\mv 0$, it thus follows that $\text{rank}(\mv S^{\star}) \leq 1$. Thus, Proposition \ref{rankone} is proved.

\vspace{-10pt}\subsection{Proof of Lemma \ref{structure}}\label{structureproof}
Without loss of generality, we denote $\mv q=a\hat{\mv q}$ with $a\geq 0$ and $\Vert\hat{\mv q}\Vert=1$. Accordingly, problem (P5) can be re-expressed as
\begin{align}
\text{(P5.2):}~\max\limits_{\hat{p},\hat{\bm q},a\geq 0}~&\frac{\hat{p}}{H_{\min}^{2}+a^{2}}\\
\text{s.t.}~&\frac{\hat{\beta}_{0}\hat{p}}{H_{\min}^{2}+\Vert a\hat{\mv q}-\mv w_{1}\Vert^{2}}\leq \hat{\Gamma},\\
&\Vert\hat{\mv q}\Vert=1,\\
&\text{(\ref{staticpower}).}\nonumber
\end{align}

Under any given feasible $\hat{p}$, optimizing $a$ and $\hat{\mv q}$ in (P5.2) is equivalent to solving
\begin{align}
\text{(P5.3):}~\min\limits_{\hat{\bm q},a\geq 0}~&a\nonumber\\
\text{s.t.}~&\Vert a\hat{\mv q}-\mv w_{1}\Vert^{2}\geq \frac{\hat{\beta}_{0}\hat{p}}{\hat{\Gamma}}-H_{\min}^{2},\label{proofq}\\
&\Vert\hat{\mv q}\Vert=1.
\end{align}
On one hand, if $\sqrt{\frac{\hat{\beta}_{0}\hat{p}}{\hat{\Gamma}}-H_{\min}^{2}}\leq \Vert\mv w_{1}\Vert$, then it is easy to verify that $a=0$ is the optimal solution to (P5.3). Thus, Lemma \ref{structure} directly follows. On the other hand, if $\sqrt{\frac{\hat{\beta}_{0}\hat{p}}{\hat{\Gamma}}-H_{\min}^{2}}> \Vert\mv w_{1}\Vert$, then $a=0$ becomes infeasible and constraint (\ref{proofq}) should be tight. In this case, constraint (\ref{proofq}) can be rewritten as $a^{2}\Vert\hat{\mv q}-\mv w_{1}/a\Vert^{2}=\hat{\beta}_{0}\hat{p}/\hat{\Gamma}-H_{\min}^{2}$. As a consequence, $a$ is minimized only when $\hat{\mv q}$ is chosen such that $\Vert\hat{\mv q}-\mv w_{1}/a\Vert$ is maximized. As a result, $\hat{\mv q}=-\frac{\mv w_{1}}{\Vert\mv w_{1}\Vert}$ must hold. By substituting $\hat{\mv q}=-\frac{\mv w_{1}}{\Vert\mv w_{1}\Vert}$ into $\mv q=a\hat{\mv q}$, we have $\mv q=-a\frac{\mv w_{1}}{\Vert\mv w_{1}\Vert}$ with $a>0$. By combining the two cases, Lemma \ref{structure} is proved.

\vspace{-10pt}\subsection{Proof of Proposition \ref{1PRhover2}} \label{1PRhover2proof}
First, we consider a relaxed problem of (P5.1) with the maximum transmit power constraint $\hat{p}\leq \hat{p}$ ignored, denoted as (P5.4). It is evident that with the optimal solution to (P5.4), the IT constraint (\ref{IT1PR1}) must be tight, and thus we have
\begin{align}\label{tildep1}
\tilde{p}=\frac{\hat{\Gamma}}{\hat{\beta}_{0}}\left((a+\Vert\mv w_{1}\Vert)^{2}+H_{\min}^{2}\right).
\end{align}
By substituting the above into the objective function of (P5.4), it can be recast as $\phi(a)=((a+\Vert\mv w_{1}\Vert)^{2}+H_{\min}^{2})/(a^{2}+H_{\min}^{2}).$ By checking the first-order derivative of $\phi(a)$ with respect to $a$, the optimal $a$ can be obtained as $\tilde{a}^{\star}$, as given in (\ref{optimala1}). By substituting this into (\ref{tildep1}), we have $\tilde{p}^{\star}$ given in (\ref{optimalp1}).

Next, we consider the problem (P5.1) with the maximum transmit power constraint $\hat{p}\leq\hat{p}$ considered. We prove this proposition by considering the following three cases, respectively.

If $\hat{p}>\tilde{p}^{\star}$, the optimal solution to problem (P5.4) is also feasible to problem (P5.1). As the objective value of (P5.4) serves as an upper bound on that of (P5.1), it follows that such a solution is also optimal to (P5.1).

If $\hat{p}<p_{1}$, it is evident that the UAV should hover exactly above the SR at the minimum altitude and transmit at the maximum power to maximize the received power at the SR, with the PR's IT constraint satisfied. Therefore, we have $\hat{p}^{\star}=\hat{p}$ and $a^{\star}=0$ in this case.

If $p_{1}\leq \hat{p}\leq \tilde{p}^{\star}$, the IT constraint (\ref{IT1PR1}) must be tight at the optimality, since otherwise we can always move the UAV closer to the SR and/or increase the UAV's transmit power to increase the SR's achievable rate, without violating the PR's IT constraint. Therefore, we have $\hat{p}=\tilde{p}$ in (\ref{tildep1}), or equivalently, $a=\sqrt{\hat{\beta}_{0}\hat{p}/\hat{\Gamma}-H_{\min}^{2}}-\Vert\mv w_{1}\Vert$. Given this, the objective function of (P5.1) can be recast as $\hat{\phi}(\hat{p})=\hat{p}/(\hat{\beta}_{0}\hat{p}/\hat{\Gamma}+\Vert\mv w_{1}\Vert^{2}-2\Vert\mv w_{1}\Vert\sqrt{\hat{\beta}_{0}\hat{p}/\hat{\Gamma}-H_{\min}^{2}})$. It is easy to verify that $\hat{\phi}(\hat{p})$ monotonically increases with $\hat{p}\in[0,\hat{p}]$. Thus, we have $\hat{p}^{\star}=\hat{p}$ and $a^{\star}=\sqrt{\hat{\beta}_{0}\hat{p}/\hat{\Gamma}-H_{\min}^{2}}-\Vert\mv w_{1}\Vert$.

By combining the above three cases, Proposition \ref{1PRhover2} is proved.

\end{document}